\documentclass[12pt]{article}%
\usepackage{amsmath}
\usepackage{amsfonts}
\usepackage{amssymb}
\usepackage{graphicx}%
\begin{document}

\begin{titlepage}
\begin{center}
\renewcommand{\thefootnote}{\fnsymbol{footnote}}
{\Large\bf  Applications of the Gauge Principle to Gravitational
Interactions} \vskip20mm {\large\bf{Ali H.
Chamseddine \footnote{email: chams@aub.edu.lb}}}\\
\renewcommand{\thefootnote}{\arabic{footnote}}
\vskip2cm {\it Center for Advanced Mathematical Sciences (CAMS)
and\\
Physics Department, American University of Beirut, Lebanon.\\}
\end{center}
\begin{center}
{\bf Abstract}
\end{center}
The idea of applying the gauge principle to formulate the general
theory of relativity started with Utiyama in 1956. I review various
applications of the gauge principle applied to different aspects of
the gravitational interactions.
\end{titlepage}
\tableofcontents

\section{\bigskip Introduction}

Gauge invariance now plays a fundamental role in theoretical physics. In order
to have a Lorentz invariant formulation of electromagnetism, one is forced to
represent the two polarizations of the massless photon with a four-component
vector. The redundancy present in the additional components of the four-vector
is then reflected by the fact that the Lagrangian is invariant under an
additional gauge transformation that allows to fix one of the space components
of the vector $A_{i}$, the other component $A_{0\text{ }}$being non dynamcial.
In general relativity the problem is more difficult as one must represent the
two polarizations of the massless graviton by a symmetric four-dimensional
tensor with ten independent components. What makes this possible is the
diffeomorphism invariance of the action which allows to fix four of the ten
parameters, the other four components $g_{0i}$ being non-dynamical \cite{ADM}.
This, however, is not the unique way to represent the gravitational field. The
other possibility, originally due to Weyl \cite{Weyl} and Cartan
\cite{Cartan}, and in a more concrete form to Utiyama \cite{Utiyama} and
Kibble \cite{Kibble}, is based on representing the two polarizations of the
graviton by the vierbein $e_{\mu}^{a}$ with sixteen independent components
which transform as a four-vector under diffeomorphisms and a Lorentz vector
under Lorentz transformations. The action is constructed to be invariant under
both diffeomorphism and local Lorentz transformations. The gauge field
associated with local Lorentz invariance, $\omega_{\mu}^{\;ab}$ is not taken
to be an independent field, but rather determined by setting the generalized
gauge covariant derivative of the field $e_{\mu}^{a}$, with respect to both
symmetries, to vanish \cite{van}%
\[
\mathcal{D}_{\mu}e_{\nu}^{a}=\partial_{\mu}e_{\nu}^{a}+\omega_{\mu}%
^{\;ab}e_{\nu}^{b}-\Gamma_{\mu\nu}^{\rho}e_{\rho}^{a}=0,
\]
where $\Gamma_{\mu\nu}^{\rho}=\Gamma_{\nu\mu}^{\rho}$ is a symmetric
connection. This system of 64 independent equations allows to solve uniquely
for the 24 components of the spin-connection $\omega_{\mu}^{\;ab}$ and for the
40 independent components $\Gamma_{\mu\nu}^{\rho}$ in function of $e_{\mu}%
^{a}$ provided that $e_{\mu}^{a}$ is invertible%
\begin{align*}
\omega_{\mu ab}\left(  e\right)   &  =\frac{1}{2}e_{a}^{\nu}e_{b}^{\rho
}\left(  \Omega_{\mu\nu\rho}\left(  e\right)  -\Omega_{\nu\rho\mu}\left(
e\right)  +\Omega_{\rho\mu\nu}\left(  e\right)  \right)  ,\\
\Omega_{\mu\nu\rho}  &  =\left(  \partial_{\mu}e_{\nu}^{c}-\partial_{\nu
}e_{\mu}^{c}\right)  e_{\rho c},\\
\Gamma_{\mu\nu}^{\rho}  &  =\frac{1}{2}g^{\rho\sigma}\left(  g_{\mu\sigma,\nu
}+g_{\nu\sigma,\mu}-g_{\mu\nu,\sigma}\right)  ,\\
g_{\mu\nu}  &  =e_{\mu}^{a}e_{\nu a},
\end{align*}
which shows that the connection $\Gamma_{\mu\nu}^{\rho}$ is identical to the
Christoffel connection. Taking the antisymmetrized derivative of the metric
condition gives an identity between the curvature of the spin-connection and
the curvature of the Christoffel connection
\begin{align*}
0  &  =R_{\mu\nu}^{\quad ab}\left(  \omega\right)  e_{\rho b}-R_{\;\rho\mu\nu
}^{\sigma}\left(  \Gamma\right)  e_{\sigma}^{a},\\
R_{\mu\nu}^{\quad ab}\left(  \omega\right)   &  =\partial_{\mu}\omega_{\nu
}^{\;ab}-\partial_{\nu}\omega_{\mu}^{\;ab}+\omega_{\mu}^{\;ac}\omega_{\nu
c}^{\;\;b}-\omega_{\nu}^{\;ac}\omega_{\mu c}^{\;\;b},\\
R_{\;\rho\mu\nu}^{\sigma}\left(  \Gamma\right)   &  =\partial_{\mu}\Gamma
_{\nu\rho}^{\sigma}-\partial_{\nu}\Gamma_{\mu\rho}^{\sigma}-\Gamma_{\mu\rho
}^{\lambda}\Gamma_{\nu\lambda}^{\sigma}+\Gamma_{\nu\rho}^{\lambda}\Gamma
_{\mu\lambda}^{\sigma},
\end{align*}
This identity implies that an invariant Lagrangian can be formed out of either
curvature
\[
e_{a}^{\mu}e_{b}^{\nu}R_{\mu\nu}^{\quad ab}\left(  \omega\right)  =g^{\rho\nu
}R_{\;\rho\mu\nu}^{\mu}\left(  \Gamma\right)  \equiv R.
\]
In this way, the Lorentz and diffeomorphism invariant Lagrangian becomes
either a function of $e_{\mu}^{a}$ only or a function of $g_{\mu\nu}$ only.
The equivalence of both expressions can be seen by noting that the Lorentz
gauge transformations can be used to fix the six antisymmetric components of
$e_{\mu}^{a}$ to vanish. Then in both cases diffeomorphism invariance fixes
four more components out of the six $g_{ij}$,$i,j=1,2,3,$ with the four
components $g_{0i}$ being non dynamical, leaving only two dynamical degrees of
freedom, as should be the case. This shows the equivalence of the vierbein and
metric formulations of the general theory of relativity, a result which has
been dealt with intensively in the literature. This classical equivalence was
culminated in the work of Ashtekar \cite{ashtekhar}, and collaborators where
it was shown that the $SL(2,\mathbb{C})$ invariance can be written in such a
way as to have a manifest invariance under the complex group $SU(2)$
\ \cite{sen} so that the gauge fields could be separated into self-dual and
anti self-dual parts. The main advantage of the Ashtekar formulation is that
it allows to take the spin-connection as the canonical variables while the
inverse of the soldering forms $e_{a}^{\mu}$ are taken as the conjugate
variables, which allowed to formulate a theory for loop quantum gravity
\cite{Smolin}.

The requirement that the Lagrangian should have local $SL(2,\mathbb{C})$
invariance is powerful \cite{salam} and fits with the fact that space-time
spinors do exist in nature. The Dirac equation in Minkowski space-time has
global $SL(2,\mathbb{C})$ invariance, and it is natural to require that this
invariance be promoted to become local by introducing the spin-connection as a
gauge field. There is, however, a need to introduce the vierbein, or soldering
form as an external field. Naturally, the idea of extending the homogeneous
Lorentz invariance to the inhomegeneous Lorentz invariance was exploited
\cite{Ivan}, \cite{Hayashi}, \cite{HayaBerg}. In this case the group has
translation generators in addition to the rotation generators. The field
strengths associated with the translation generators are constrained to be
zero, allowing the identification of the translation gauge parameters with the
diffeomorphisms parameters. This is where a gauge theory of gravity differs
from the usual Yang-Mills type gauge theory, as the constraint of vanishing
translational field strength allows to solve for the spin connection as
function of the vierbein, which in this case is the gauge field associated
with translations. This makes the field strength associated with rotations
depend on second derivatives of the vierbein, and becomes identified with the
curvature of the metric formed from the vierbein, as shown above. The
constraints render the theory non-renormalizable. This explains why the method
of formulating gravity as a gauge theory of the inhomogeneous Lorentz group
does not lead to improvements in the renormalizability of the theory. The main
advantages lie in the simplicity of formulation, in having a polynomial
structure, and in the prospect of extending the gravitational theory to be
unified with the other interactions.

The gauge principle has played a prominent role in the formulation of
different aspects of gravitational theories. This was one of the guiding
principles of my research for the last thirty years. In this article I\ will
review the various ways of formulating theories of gravity based on the gauge
principle. The recent interest in models of bigravity resulting from brane
models makes it necessary to study the mechanism of generating a small mass to
one combination of the two metrics in a consistent way. For some time it was
thought that a theory for a massive graviton is inconsistent because it does
not have a smooth limit to the massless case \cite{vv},\cite{za},\cite{bd}. We
shall show that, in analogy with gauge theory for spin-1 fields, it is
possible to have a consistent theory by generating the mass through
spontaneous symmetry breaking and the use of the Higgs mechanism. I\ will also
study the possibility of extending the general theory of relativity to
describe a complex Hermitian metric, an idea first considered by Einstein in
1945 \cite{Ein1},\cite{Einstein}, in his attempt to unify gravity with
electromagnetism. It is now known that the antisymmetric part of the metric
does not describe the electromagnetic field but rather an antisymmetric tensor
\cite{Moffat}, \cite{Damour1} , which for consistency at the non-linear level,
should be massive \cite{Damour2}. The other possibility to be explored is the
unification of internal symmetries and space-time symmetries. This corresponds
to a theory of a complex $U(N)$ valued metric \cite{ISS}, \cite{chams1}. We
shall also show that it is possible to construct simple supergravity
\cite{Ferrara} using the gauge method by considering graded algebras
\cite{CW}, \cite{mcdowel},\cite{ortho}, \cite{chams3}, \cite{Townsend}.
Finally, we show that when space-time coordinates do not commute, which makes
it necessary to replace ordinary products with Moyal products \cite{H}, it is
possible to deform Einstein's gravity using the methods learned from the gauge
formulations of gravity. The plan of this paper is as follows. In section two
we briefly review the formulation of gravity as a gauge theory of the
inhomogeneous Lorentz group. In section three we consider a gauge theory with
two gravitons and show how to construct a consistent theory of one massless
and one massive graviton by employing the Higgs mechanism. In section four we
consider a gauge theory of gravity based on gauging the unitary group $U(1,3)$
and show that this is closely related to general relativity with Hermitian
metric. In section five we consider unification of space-time and internal
symmetries, and show that it is possible to construct a gauge theory of
gravity with matrix valued metrics, where only one graviton remains massless
with the other gravitons acquiring mass through the spontaneous breakdown of
the larger symmetry. \ In section six we briefly discuss gauging graded
algebras to construct simple supergravity. In section seven we consider
topological gauge theories of gravity based on Chern-Simons forms with the
gauge groups taken to be generalizations of the rotation groups in higher
dimensions. In section eight we consider the noncommutative extension of the
gauge formulation of gravity where usual products are replaced with Moyal
products. Section nine is the conclusion.

\section{Gauging the inhomogeneous Lorentz group}

It is natural to generalize global invariance of the Dirac Lagrangian to a
local one under Lorentz transformations
\[
\delta\psi=\Omega\psi,
\]
where $\Omega=\exp\left(  \frac{1}{4}\lambda^{ab}\gamma_{ab}\right)  $ and
$\gamma_{ab}=\frac{1}{2}\left[  \gamma_{a},\gamma_{b}\right]  $ by allowing
the gauge parameters $\lambda^{ab}$ to depend on $x^{\mu}.$ The invariance is
achieved by replacing the ordinary derivative $\partial_{\mu}$ with the
covariant derivative \cite{Utiyama}, \cite{Kibble}
\[
\nabla_{\mu}=\partial_{\mu}+\frac{1}{4}\omega_{\mu}^{\;ab}\gamma_{ab}.
\]
Utiyama proposed to realize gravity as a gauge theory of the homogeneous
Lorentz group with local gauge invariance under space-time coordinate
rotations. This step was then generalized to the inhomegeneous Lorentz group
to enforce invariance under local translations as well. The connection
associated with this group is given by \cite{Ivan}, \cite{Hayashi}%
,\cite{HayaBerg},\cite{cho}
\[
\nabla_{\mu}=\partial_{\mu}+e_{\mu}^{a}P_{a}+\omega_{\mu}^{\;ab}J_{ab},
\]
where $P_{a}$ are the group generators associated with translation and
$J_{ab}$ are the rotation generators, with $e_{\mu}^{a}$ and $\omega_{\mu
}^{\;ab}$ being the respective gauge fields. The curvature associated with
this connection is evaluated to give
\[
\left[  \nabla_{\mu},\nabla_{\nu}\right]  =T_{\mu\nu}^{\quad a}P_{a}+R_{\mu
\nu}^{\;\;\;ab}J_{ab},
\]
where
\begin{align*}
T_{\mu\nu}^{\quad a} &  =\partial_{\mu}e_{\nu}^{a}+\omega_{\mu}^{\;ab}e_{\nu
b}-\mu\leftrightarrow\nu,\\
R_{\mu\nu}^{\;\;\;ab} &  =\partial_{\mu}\omega_{\nu}^{\;ab}+\omega_{\mu
}^{\;ac}\omega_{\nu c}^{\quad b}-\mu\leftrightarrow\nu.
\end{align*}
The gauge transformation of the vierbein $e_{\mu}^{a}$ is given by%
\[
\delta e_{\mu}^{a}=\partial_{\mu}\zeta^{a}+\omega_{\mu}^{\quad ab}\zeta
_{b}+\lambda^{ab}e_{\mu b}.
\]
Setting the torsion $T_{\mu\nu}^{\quad a}$ to zero allows to solve for
$\omega_{\mu}^{\;ab}$ uniquely provided that the field $e_{\mu}^{a}$ is
invertible. This is the same equation as that obtained by antisymmetrizing the
metric condition%
\[
\mathcal{D}_{\mu}e_{\nu}^{a}-\mathcal{D}_{\nu}e_{\mu}^{a}=T_{\mu\nu}%
^{\;\;a}=0.
\]
The number of independent components in $T_{\mu\nu}^{\;\;a}$ matches the
number of independent components in $\omega_{\mu}^{\;ab}.$ The presence of
this constraint is the main difference between gravity and Yang-Mills gauge
theories. This constraint breaks the translational gauge invariance. We note,
however, that we can write for the gauge transformation of $e_{\mu}^{a},$
\begin{align*}
\delta e_{\mu}^{a} &  =\partial_{\mu}\zeta^{\nu}e_{\nu}^{a}+\zeta^{\nu
}\partial_{\mu}e_{\nu}^{a}+\omega_{\mu}^{\;ab}\zeta_{b}+\lambda^{ab}e_{\mu
b}\\
&  =\partial_{\mu}\zeta^{\nu}e_{\nu}^{a}+\zeta^{\nu}\left(  \partial_{\nu
}e_{\mu}^{a}+\omega_{\nu}^{\;ab}e_{\mu b}-\omega_{\mu}^{\;ab}e_{\nu b}%
+T_{\mu\nu}^{\;\;a}\right)  +\omega_{\mu}^{\;ab}\zeta_{b}+\lambda^{ab}e_{\mu
b}\\
&  =\partial_{\mu}\zeta^{\nu}e_{\nu}^{a}+\zeta^{\nu}\partial_{\nu}e_{\mu}%
^{a}+\lambda^{^{\prime}ab}e_{\mu b}+\zeta^{\nu}T_{\mu\nu}^{\;\;a},
\end{align*}
where $\zeta^{\nu}=\zeta^{a}e_{a}^{\nu}$ becomes the parameter for general
coordinate transformations provided that the torsion vanishes. One can require
that the transformations of $\omega_{\mu}^{\;ab}$ be modified so as to
preserve the zero torsion constraint. The action is constructed in terms of
the gauge covariant curvature $R_{\mu\nu}^{\;\;ab}$ and the gauge fields
$e_{\mu}^{a}$%
\[
I=%
{\displaystyle\int}
d^{4}x\epsilon^{\mu\nu\kappa\lambda}\epsilon_{abcd}e_{\mu}^{a}e_{\nu}%
^{b}R_{\kappa\lambda}^{\;\;cd}\left(  \omega\right)  .
\]
This is invariant under Lorentz rotations and diffeomorphisms. It is also
invariant under translation invariance provided the torsion is set to zero:
\begin{align*}
\delta I &  =2%
{\displaystyle\int}
d^{4}x\epsilon^{\mu\nu\kappa\lambda}\epsilon_{abcd}\nabla_{\mu}\zeta^{a}%
e_{\nu}^{b}R_{\kappa\lambda}^{\;\;\;cd}\left(  \omega\right)  \\
&  =-%
{\displaystyle\int}
d^{4}x\epsilon^{\mu\nu\kappa\lambda}\epsilon_{abcd}\zeta^{a}T_{\mu\nu}%
^{b}R_{\kappa\lambda}^{\;\;\;cd}\left(  \omega\right)  \\
&  =0,
\end{align*}
where we have integrated by parts and used the Bianchi identity $\nabla
_{\left[  \mu\right.  }R_{\left.  \kappa\lambda\right]  }^{\;\;\;\;cd}\left(
\omega\right)  =0.$ Therefore enforcing the zero torsion constraint allows to
consider an invariant action, without the need to define a metric on the
four-dimensional manifold, by making use of the differential form
representation. Thus, let \cite{ISS2}
\[
e^{a}=e_{\mu}^{a}dx^{\mu},\quad\omega^{ab}=\omega_{\mu}^{\;ab}dx^{\mu},
\]
and
\[
R^{ab}=d\omega^{ab}+\omega^{ac}\wedge\omega_{c}^{\;b}\equiv\frac{1}{2}%
R_{\mu\nu}^{\;\;\;ab}dx^{\mu}\wedge dx^{\nu},
\]
so that the invariant action simplifies to
\[
I=%
{\displaystyle\int\limits_{M}}
\epsilon_{abcd}e^{a}\wedge e^{b}\wedge R^{cd}.
\]
To this one can always add a cosmological constant
\[%
{\displaystyle\int\limits_{M}}
\epsilon_{abcd}e^{a}\wedge e^{b}\wedge e^{c}\wedge e^{d}.
\]
One can further write the above action in an index free notation be utilizing
the $SL(2,\mathbb{C})$ invariance \cite{salam}. Thus, let
\begin{align*}
e &  =e^{a}\gamma_{a},\\
\omega &  =\frac{1}{4}\omega^{ab}\gamma_{ab},
\end{align*}
which transform as
\[
e\rightarrow\Omega^{-1}e\,\Omega,\quad\omega\rightarrow\Omega^{-1}%
\omega\,\Omega+\Omega^{-1}d\,\Omega,
\]
under $SL(2,\mathbb{C})$ transformations. The action then simplifies to
\[
I=2%
{\displaystyle\int\limits_{M}}
Tr\left(  i\gamma_{5}e\wedge e\wedge R\right)  ,
\]
where
\[
R=d\omega+\omega\wedge\omega,
\]
and invariance of the action follows from the commutativity of $\gamma_{5}$
with $\Omega$ \cite{ncactions}.

\section{Massive gravity through spontaneous symmetry breaking}

In the Kaluza-Klein approach of gravity in higher dimensions, the components
of the higher dimensional metric tensor along the four-dimensional subspace is
expanded in terms of Fourier components
\[
g_{\mu\nu}\left(  x,y\right)  =%
{\displaystyle\prod\limits_{i=1}^{D-4}}
{\displaystyle\sum\limits_{n_{i}=0}^{\infty}}
g_{\mu\nu n_{1}\cdots n_{D-4}}\left(  x\right)  e^{in_{i}y^{i}}%
\]
where $y^{i}$ are the coordinates of the compact directions. Besides the zero
mode representing the massless graviton, one gets an infinite number of
massive gravitons whose masses which are multiples of the Planck mass. In
brane models of gravity as well in gravitational models in noncommutative
geometry it is possible to get gravitons with small mass \cite{dk}. The
linearized Lagrangian for massive spin-2 field $h_{\mu\nu}$ was found by Fierz
and Pauli to be given by \cite{fp}
\begin{align*}
I  &  =-\frac{1}{4}%
{\displaystyle\int}
d^{4}x\left(  \partial_{\lambda}h_{\mu\nu}\partial^{\lambda}h^{\mu\nu
}-2\partial^{\nu}h_{\mu\nu}\partial_{\lambda}h^{\mu\lambda}+2\partial^{\nu
}h_{\mu\nu}\partial^{\mu}h_{\lambda}^{\;\lambda}-\partial_{\mu}h_{\nu
}^{\,\;\nu}\partial^{\mu}h_{\lambda}^{\;\lambda}\right. \\
&  \hspace{0.7in}\left.  +m^{2}\left(  h_{\mu\nu}h^{\mu\nu}-bh_{\nu}^{\,\;\nu
}h_{\lambda}^{\;\lambda}\right)  \right)  ,
\end{align*}
where, for consistency $b$ must be set to 1 so as to guarantee that only five
components of $h_{\mu\nu}$ propagate, instead of the expected six. The mass
independent part of the above Lagrangian is the same as the one obtained by
linearizing the Einstein-Hilbert action around a Minkowski background. The
propagator for $h_{\mu\nu}$ is \cite{salam2}
\begin{align*}
\Delta_{\mu\nu}^{\rho\sigma}  &  =\frac{1}{m^{2}-k^{2}}\left(  \left(
\delta_{\mu}^{\rho}-\frac{k_{\mu}k^{\rho}}{m^{2}}\right)  \left(  \delta_{\nu
}^{\sigma}-\frac{k_{\nu}k^{\sigma}}{m^{2}}\right)  \right. \\
&  \hspace{0.5in}-\frac{1}{3}\left(  \eta_{\mu\nu}-\frac{k_{\mu}k_{\nu}}%
{m^{2}}\right)  \left(  \eta^{\rho\sigma}-\frac{k^{\rho}k^{\sigma}}{m^{2}%
}\right) \\
&  \hspace{0.5in}\left.  +\frac{1-b}{2\left(  1-b\right)  k^{2}+\left(
1-4b\right)  m^{2}}\left(  \eta_{\mu\nu}+\frac{2k_{\mu}k_{\nu}}{m^{2}}\right)
\left(  \eta^{\rho\sigma}+\frac{2k^{\rho}k^{\sigma}}{m^{2}}\right)  \right)  .
\end{align*}
Notice that when $b\neq1$, the massive spin-2 field and the ghost of the
spin-0 are coupled, and will only decouple for $b=1,$ which is the Fierz-Pauli
choice. The choice $b=1$ cannot be maintained at the quantum level and has to
be tuned. The symmetric tensor $h_{\mu\nu}$ has ten independent components but
only the six components $h_{ij}$ $(i,j=1,2,3)$ are dynamical. A massive spin-2
field must have only five dynamical degrees of freedom $(2j+1=5)$. This
implies that there is an additional component, a spin-0 ghost, that does not
decouple except for the choice $b=1.$ The limit of this propagator to the
massless case $m\rightarrow0$ is singular and is similar to the propagator of
a massive spin-1 field, which is also singular in the massless limit. This
strongly suggests that in order to solve the problem of the singular zero mass
limit, the mass of the spin-2 field should be acquired through spontaneous
symmetry breaking and the Higgs mechanism. To achieve this, the Lagrangian
must have a gauge symmetry to be broken. This makes it necessary to extend the
symmetry of the system, but in such a way as not to increase the dynamical
degrees of freedom of the system. This is where the gauge principle of
formulating gravity enters.

To illustrate the mechanism, we consider a {coupled system of one massless
graviton and one massive graviton formulated as a gauge theory of $SP(4)\times
SP(4)$ \cite{css}. }We start with the gauge fields
\begin{align*}
A_{\mu\alpha}^{\hspace{0.12in}\beta}  &  =\left(  ie_{\mu}^{a}\gamma_{a}%
+\frac{1}{4}\omega_{\mu}^{\;ab}\gamma_{ab}\right)  _{\alpha}^{\beta},\\
A_{\mu\alpha}^{^{\prime\hspace{0.12in}\beta}}  &  =\left(  ie_{\mu}^{^{\prime
}a}\gamma_{a}+\frac{1}{4}\omega_{\mu}^{^{\prime}\;ab}\gamma_{ab}\right)
_{\alpha}^{\beta},
\end{align*}
where $\left(  A_{\mu}C\right)  _{\alpha\beta}=\left(  A_{\mu}C\right)
_{\beta\alpha}$, $C$ \ being the charge conjugation matrix. These also satisfy
the reality conditions
\[
\gamma_{0}A_{\mu}^{\dagger}\gamma_{0}=-A_{\mu},\quad\gamma_{0}A_{\mu
}^{^{\prime}\dagger}\gamma_{0}=-A_{\mu}^{^{\prime}}.
\]
\ These have the following gauge transformations
\begin{align*}
A_{\mu}  &  \rightarrow\Omega A_{\mu}\Omega^{-1}+\Omega\partial_{\mu}%
\Omega^{-1}\\
A_{\mu}^{^{\prime}}  &  \rightarrow\Omega^{^{\prime}}A_{\mu}^{^{\prime}}%
\Omega^{^{\prime}-1}+\Omega^{^{\prime}}\partial_{\mu}\Omega^{^{\prime}-1}%
\end{align*}
where $\Omega$ and $\Omega^{^{\prime}}$ denote independent symplectic matrices
as gauge parameters%
\begin{align*}
\Omega &  =\exp\left(  i\lambda^{a}\gamma_{a}+\frac{1}{4}\lambda^{ab}%
\gamma_{ab}\right)  ,\\
\Omega^{^{\prime}}  &  =\exp\left(  i\lambda^{^{\prime}a}\gamma_{a}+\frac
{1}{4}\lambda^{^{\prime}ab}\gamma_{ab}\right)  ,
\end{align*}
satisfying $\Omega^{-1}=C\Omega^{T}C^{-1}.$ Notice that the field $e_{\mu}%
^{a}$ is now associated with translation in the internal space $SP(4)$. In
other words, the $SL(2,\mathbb{C})$ symmetry is now extended to $SP(4)$. Next,
introduce a Higgs fields $G$ subject to the reality condition%
\[
\gamma_{0}G^{\dagger}\gamma_{0}=CG^{T}C^{-1}\equiv\widetilde{G}%
\]
transforming under the product representation of {$SP(4)\times SP(4)$}%
\[
G\rightarrow\Omega G\Omega^{^{\prime}-1},
\]
which has the following expansion in the Clifford algebra basis%
\[
G_{\alpha}^{\beta^{^{\prime}}}=\left(  \varphi+i\pi\gamma_{5}+iv^{a}\gamma
_{5}\gamma_{a}+ig^{a}\gamma_{a}+g^{ab}\gamma_{ab}\right)  _{\alpha}%
^{\beta^{^{\prime}}}.
\]
For the theory to possess a stable Poincar\'{e} invariant vacuum solution, we
require $G$ to have a non-vanishing vacuum:%
\[
\left\langle G_{\alpha}^{\beta^{^{\prime}}}\right\rangle =\left(
a+ib\gamma_{5}\right)  _{\alpha}^{\beta^{^{\prime}}}.
\]
This breaks the symmetry spontaneously from {$SP(4)\times SP(4)$ to
$SL(2,\mathbb{C})$ through a non-linear realization \cite{CWZ}. The number of
independent components of }$G$ needed {to parametrize the homogeneous space }%
\[
\frac{{SP(4)\times SP(4)}}{{SL(2,\mathbb{C})}},
\]
is $10+10-6=14$. Thus two constraints must be imposed on $G$ to reduce the
number of independent components from $16$ to $14.$ For example these can be
taken to be
\begin{align*}
Tr\left(  G\widetilde{G}\right)   &  =4c_{1},\\
Tr\left(  \left(  G\widetilde{G}\right)  ^{2}\right)   &  =4c_{2}%
,\hspace{0.5in}%
\end{align*}
where $\widetilde{G}=CG^{T}C^{-1}\rightarrow\Omega^{^{\prime}}\widetilde
{G}\Omega^{-1}$. The action is taken to be of the form{%
\begin{align*}
&  \int Tr\left(  \alpha G\widetilde{G}F\wedge F+\alpha^{^{\prime}}%
\widetilde{G}GF^{^{\prime}}\wedge F^{^{\prime}}\right. \\
&  \hspace{0.3in}\left.  +\beta\nabla G\wedge\nabla\widetilde{G}\wedge\nabla
G\wedge\nabla\widetilde{G}+\beta^{^{\prime}}\nabla G\wedge\nabla\widetilde
{G}\wedge\nabla G\wedge\nabla\widetilde{G}\right)  ,
\end{align*}
}where
\begin{align*}
F  &  =dA+A\wedge A,\\
F^{^{\prime}}  &  =dA^{^{\prime}}+A^{^{\prime}}\wedge A^{^{\prime}},\\
\nabla G  &  =dG+AG-GA^{^{\prime}},\\
\nabla\widetilde{G}  &  =d\widetilde{G}+A^{^{\prime}}\widetilde{G}%
-\widetilde{G}A,
\end{align*}

To analyze the physical content of the Lagrangian we can eliminate the $14$
remaining components of $G$ by fixing $14$ gauge conditions, with the
remaining gauge freedom corresponding to the unbroken local ${SL(2,\mathbb{C}%
)}$ invariance. To derive the component form of the Lagrangian in this unitary
gauge, we write%
\begin{align*}
F &  =\frac{1}{2}\left(  iF_{\mu\nu}^{\;\;\;a}\gamma_{a}+F_{\mu\nu}%
^{\;\;\;ab}\gamma_{ab}\right)  dx^{\mu}\wedge dx^{\nu}\\
F^{^{\prime}} &  =\frac{1}{2}\left(  iF_{\mu\nu}^{^{\prime}\;\;\;a}\gamma
_{a}+F_{\mu\nu}^{^{\prime}\;\;\;ab}\gamma_{ab}\right)  dx^{\mu}\wedge dx^{\nu
}\\
F_{\mu\nu}^{\quad a} &  =T_{\mu\nu}^{\;\;\;a},\qquad F_{\mu\nu}^{^{\prime
}\quad a}=T_{\mu\nu}^{^{\prime}\;\;a}\\
F_{\mu\nu}^{\;\;\;ab} &  =R_{\mu\nu}^{\;\;\;ab}-4\left(  e_{\mu}^{a}e_{\nu
}^{b}-e_{\nu}^{a}e_{\mu}^{b}\right)  ,\qquad F_{\mu\nu}^{^{\prime}%
\;\;ab}=R_{\mu\nu}^{^{\prime}\;ab}-4\left(  e_{\mu}^{^{\prime}a}e_{\nu
}^{\prime b}-e_{\nu}^{^{\prime}a}e_{\mu}^{^{\prime}b}\right)  \\
\nabla_{\mu}G &  =\left(  i\left(  ae_{\mu}^{a-}-ibe_{\mu}^{a+}\gamma
_{5}\right)  \gamma_{a}+\frac{1}{4}\omega_{\mu}^{\;-ab}\left(  a+ib\gamma
_{5}\right)  \gamma_{ab}\right)  \\
\nabla\widetilde{_{\mu}G} &  =\left(  -i\left(  ae_{\mu}^{a-}+ibe_{\mu}%
^{a+}\gamma_{5}\right)  \gamma_{a}-\frac{1}{4}\omega_{\mu}^{\;-ab}\left(
a+ib\gamma_{5}\right)  \gamma_{ab}\right)
\end{align*}
where
\begin{align*}
e_{\mu}^{a\pm} &  =e_{\mu}^{a}\pm e_{\mu}^{^{\prime}a},\quad\\
\omega_{\mu}^{\;-ab} &  =\omega_{\mu}^{\;ab}-\omega_{\mu}^{\;^{\prime}ab}.
\end{align*}
One then finds that the action simplifies to \cite{css}
\begin{align*}
I &  =%
{\displaystyle\int\limits_{M}}
d^{4}x\epsilon^{\mu\nu\rho\sigma}\left(  \alpha_{1}F_{\mu\nu ab}F_{\rho\sigma
}^{\;\;ab}+\alpha_{1}^{\prime}F_{\mu\nu ab}^{^{\prime}}F_{\rho\sigma
}^{^{\prime}\;\;ab}\right)  \\
&  +\epsilon_{abcd}%
{\displaystyle\int\limits_{M}}
d^{4}x\epsilon^{\mu\nu\rho\sigma}\left(  \beta_{1}F_{\mu\nu}^{\;\;\;ab}%
F_{\rho\sigma}^{\;\;\;cd}+\beta_{1}^{\prime}F_{\mu\nu}^{^{\prime\;\;}%
ab}F_{\rho\sigma}^{^{\prime}\;\;cd}\right)  \\
&  +\epsilon_{abcd}%
{\displaystyle\int\limits_{M}}
d^{4}x\epsilon^{\mu\nu\rho\sigma}\left(  \gamma_{1}e_{\mu}^{a-}e_{\nu}%
^{b-}e_{\rho}^{c-}e_{\sigma}^{d+}+\gamma_{1}^{^{\prime}}e_{\mu}^{a+}e_{\nu
}^{b+}e_{\rho}^{c+}e_{\sigma}^{d-}\right)  \\
&  +\epsilon_{abcd}%
{\displaystyle\int\limits_{M}}
d^{4}x\epsilon^{\mu\nu\rho\sigma}\left(  \delta_{1}\left(  e_{\mu}^{a+}e_{\nu
e}^{-}-e_{\mu}^{a-}e_{\nu e}^{e}\right)  \omega_{\rho}^{\;-bc}\omega_{\sigma
}^{\;-de}+\delta_{1}^{^{\prime}}e_{\mu}^{a+}e_{\nu}^{b-}\omega_{\rho}%
^{\;-be}\omega_{\sigma e}^{\;-d}\right)  ,
\end{align*}
where the coefficients appearing above depend on $\alpha,\alpha^{\prime}%
,\beta,\beta^{\prime},\gamma,\gamma^{\prime},a$ and $b.$ Notice that no metric
is needed to define this action and all terms appearing correspond to
four-forms. It contains topological terms corresponding to Euler and
Gauss-Bonnet invariants, as well as kinetic terms for the two metrics $e_{\mu
}^{a+}$ and $e_{\mu}^{a-}.$ The physical spectrum of a similar Lagrangian was
carried a long time ago \cite{css}, where it was shown that one obtains one
combination of the two vierbeins to represent a massless graviton with the
other combination representing a massive graviton whose mass can be adjusted
to take very small values. The linearized form of the above Lagrangian is of
the Pauli-Fierz type where there are two dynamical degrees of freedom for the
massless graviton and five degrees for the massive graviton. As mentioned
earlier, the propagator for the massive graviton is singular in the zero mass
limit. It is then necessary to study such a limit in a non-unitary gauge,
where the extended gauge invariance is manifest, and where the components of
the Higgs field $G$ are not gauge fixed. For example one can partially fix $G$
to take the form
\[
G=a+i\pi\gamma_{5}+iv^{a}\gamma_{5}\gamma_{a}.
\]
In this case one can show that the physical degrees of freedom correspond to
each of the two massless spin-2 polarizations for $e_{\mu}^{a+}$ and $e_{\mu
}^{a-}$ as well as two degrees for the transverse polarizations in $v^{a}$ and
one degree for the spin-0 mode, which can be taken as the field $\pi$ or the
spin-0 longitudinal component of $v^{a}.$ For details see \cite{spontan}. The
instability of the Fierz-Pauli choice of the mass terms would occur at the
quantum level, but as explained in \cite{ags}, the corrections would occur at
a cut-off energy, where the ghost mode would start to propagate. At energies
much lower than the cut-off scale, the corrections could be ignored, and the
system is well behaved. We conclude that the proper way of formulating a
consistent theory for massive gravity is to adopt the gauge principle for
gravitation with extended symmetry and to use the Higgs mechanism to generate
mass by breaking the symmetry spontaneously to $SL(2,\mathbb{C}).$

\section{Gravity with Hermitian metric}

In his attempt to unify gravity with electromagnetism, Einstein proposed to
consider a Hermitian metric satisfying the property \cite{Ein1},
\cite{Einstein}
\begin{align*}
g_{\mu\nu}\left(  x\right)   &  =G_{\mu\nu}\left(  x\right)  +iB_{\mu\nu
}\left(  x\right)  ,\\
g_{\mu\nu}^{\dagger}\left(  x\right)   &  =g_{\nu\mu}\left(  x\right)  ,
\end{align*}
which implies the symmetries
\[
G_{\mu\nu}\left(  x\right)  =G_{\nu\mu}\left(  x\right)  ,\qquad B_{\mu\nu
}\left(  x\right)  =-B_{\nu\mu}\left(  x\right)  .
\]
In this picture the metric tensor of space-time $G_{\mu\nu}\left(  x\right)  $
is unified geometrically\ with the field strength $B_{\mu\nu}\left(  x\right)
$ of the electromagnetic field. \ There is some arbitrariness in the geometric
construction due to the non-uniqueness of the connection and a certain choice
was taken to remove this ambiguity.

The gauge formulation of gravity with a complex metric has the advantage that
it removes the above ambiguity and is very elegant. Assume that we start with
the $U(1,3)$ gauge fields $\omega_{\mu\;b}^{\,a}$ \cite{chamscomplex} The
$U(1,3)$ group of transformations is defined as the set of matrix
transformations leaving the quadratic form
\[
\left(  Z^{a}\right)  ^{\dagger}\eta_{b}^{a}Z^{b},
\]
invariant, where $Z^{a}$ are $4$ complex fields and
\[
\eta_{b}^{a}=diag\left(  -1,1,1,1\right)  ,
\]
with $3$ positive entries. The gauge fields $\omega_{\mu\,\,b}^{\,a}$ must
then satisfy the condition
\[
\left(  \omega_{\mu\,\,b}^{\,a}\right)  ^{\dagger}=-\eta_{c}^{b}\omega
_{\mu\,\,d}^{\,c}\eta_{a}^{d}.
\]
The curvature associated with this gauge field is
\[
R_{\mu\nu\;b}^{\;\;a}=\partial_{\mu}\omega_{\nu\,\,b}^{\,a}-\partial_{\nu
}\omega_{\mu\,\,b}^{\,a}+\omega_{\mu\,\,c}^{\,a}\omega_{\nu\,\,b}^{\,c}%
-\omega_{\nu\,\,c}^{\,a}\omega_{\mu\,\,b}^{\,c}.
\]
Under gauge transformations we have
\[
^{g}\omega_{\mu\,\,b}^{\,a}=M_{c}^{a}\omega_{\mu\,\,d}^{\,c}M_{b}^{-1d}%
-M_{c}^{a}\partial_{\mu}M_{b}^{-1c},
\]
where the matrices $M$ are subject to the condition:
\[
\left(  M_{c}^{a}\right)  ^{\dagger}\eta_{b}^{a}M_{d}^{b}=\eta_{d}^{c}.
\]
The curvature then transforms as
\[
^{g}R_{\mu\nu\;b}^{\;\;a}=M_{c}^{a}R_{\mu\nu\;d}^{\;\;c}M_{b}^{-1d}.
\]
Next we introduce the complex vierbein $e_{\mu}^{a}$ and its inverse
$e_{a}^{\mu}$ defined by
\[
e_{a}^{\nu}e_{\mu}^{a}=\delta_{\mu}^{\nu},\quad e_{\nu}^{a}e_{b}^{\nu}%
=\delta_{b}^{a},
\]
which transform as
\[
^{g}e_{\mu}^{a}=M_{b}^{a}e_{\mu}^{b},\quad^{g}e_{a}^{\mu}=\widetilde{e}%
_{b}^{\mu}M_{a}^{-1b}.
\]
It is also useful to define the complex conjugates
\[
e_{\mu a}\equiv\left(  e_{\mu}^{a}\right)  ^{\dagger},\quad e^{\mu a}%
\equiv\left(  e_{a}^{\mu}\right)  ^{\dagger}.
\]
With this, it is not difficult to see that
\[
e_{a}^{\mu}R_{\mu\nu\;b}^{\;\;a}\eta_{c}^{b}e^{\nu c},
\]
is hermitian and $U(1,3)$ invariant. The metric is defined by
\[
g_{\mu\nu}=\left(  e_{\mu}^{a}\right)  ^{\dagger}\eta_{b}^{a}e_{\nu}^{b},
\]
satisfy the property $g_{\mu\nu}^{\dagger}=g_{\nu\mu}.$ When the metric is
decomposed into its real and imaginary parts:
\[
g_{\mu\nu}=G_{\mu\nu}+iB_{\mu\nu},
\]
the hermiticity property then implies the symmetries
\[
G_{\mu\nu}=G_{\nu\mu},\quad B_{\mu\nu}=-B_{\nu\mu}.
\]
The gauge invariant Hermitian action is given by
\[
I=\int d^{4}x\sqrt{e}^{\dagger}e_{a}^{\mu}R_{\mu\nu\;b}^{\;\;a}\eta_{c}%
^{b}e^{\nu c}\sqrt{e},
\]
where $e=\det\left(  e_{\mu}^{a}\right)  .$ One goes to the second order
formalism by integrating out the spin connection and substituting its value in
terms of the vierbein. The resulting action depends only on the fields
$g_{\mu\nu}.$ It is worthwhile to stress that the above action, unlike others
proposed to describe nonsymmetric gravity \cite{Moffat} is unique, except for
the measure, and unambiguous. Similar ideas have been proposed in the past
based on gauging the groups $O(4,4)$ \cite{Green},\cite{MS} and $GL(4)$
\cite{siegel}, in relation to string duality, but the results obtained there
are different from what is presented here.

The infinitesimal gauge transformations for $e_{\mu}^{a}$ is $\delta e_{\mu
}^{a}=\Lambda_{\;b}^{a}e_{\mu}^{b},$ which can be decomposed into real and
imaginary parts by writing
\begin{align*}
e_{\mu}^{a}  &  =e_{0\mu}^{a}+ie_{1\mu}^{a},\\
\Lambda_{\;b}^{a}  &  =\Lambda_{0\;b}^{a}+i\Lambda_{1\;b}^{a}.
\end{align*}
From the gauge transformations of $e_{0\mu}^{a}$ and $e_{1\mu}^{a}$ one can
easily show that the gauge parameters $\Lambda_{0\;b}^{a}$ and $\Lambda
_{1\;b}^{a}$ can be chosen to make $e_{0\mu a}$ symmetric in $\mu$ and $a$ and
$e_{1\mu a}$ antisymmetric in $\mu$ and $a$. This is equivalent to the
statement that the Lagrangian should be completely expressible in terms of
$G_{\mu\nu}$ and $B_{\mu\nu}$ only, after eliminating $\omega_{\mu\,\,b}%
^{\,a}$ through its equations of motion. In reality we have
\begin{align*}
G_{\mu\nu}  &  =e_{0\mu}^{a}e_{0\nu}^{b}\eta_{ab}+e_{1\mu}^{a}e_{1\nu}^{b}%
\eta_{ab},\\
B_{\mu\nu}  &  =e_{0\mu}^{a}e_{1\nu}^{b}\eta_{ab}-e_{1\mu}^{a}e_{0\nu}^{b}%
\eta_{ab}.
\end{align*}
In this special gauge, where we define
\[
g_{0\mu\nu}=e_{0\mu}^{a}e_{0\nu}^{b}\eta_{ab},\quad g_{0\mu\nu}g_{0}%
^{\nu\lambda}=\delta_{\mu}^{\lambda},
\]
and use $e_{0\mu}^{a}$ to raise and lower indices, we get
\begin{align*}
B_{\mu\nu}  &  =-2e_{1\mu\nu},\\
G_{\mu\nu}  &  =g_{0\mu\nu}-\frac{1}{4}B_{\mu\kappa}B_{\lambda\nu}%
g_{0}^{\kappa\lambda}.
\end{align*}
The last formula appears in the metric of the effective action in open string
theory \cite{SW}.

We can express the Lagrangian in terms of $e_{\mu}^{a}$ only by solving the
$\omega_{\mu\,\,b}^{\,a}$ equations of motion
\begin{align*}
e_{a}^{\mu}e^{\nu b}\omega_{\nu\,\,b}^{\,c}+e_{b}^{\nu}e^{\mu c}\omega
_{\nu\,\,a}^{\,b}-e^{\mu b}e_{a}^{\nu}\omega_{\nu\,\,b}^{\,c}-e_{b}^{\mu
}e^{\nu c}\omega_{\nu\,\,a}^{\,b}  &  =\\
\frac{1}{\sqrt{G}}\partial_{\nu}\left(  \sqrt{G}\left(  e_{a}^{\nu}e^{\mu
c}-e_{a}^{\mu}e^{\nu c}\right)  \right)   &  \equiv X_{\quad a}^{\mu c}%
\end{align*}
where $X_{\quad a}^{\mu c}$ satisfy $\left(  X_{\quad a}^{\mu c}\right)
^{\dagger}=-X_{\quad c}^{\mu a}.$ One has to be very careful in working with a
nonsymmetric metric
\begin{align*}
g_{\mu\nu}  &  =e_{\mu}^{a}e_{\nu a},\quad g^{\mu\nu}=e^{\mu a}e_{\nu a}%
,\quad\\
g_{\mu\nu}g^{\nu\rho}  &  =\delta_{\mu}^{\rho},\quad g_{\mu\nu}g^{\mu\rho}%
\neq\delta_{\mu}^{\rho}.
\end{align*}
Care also should be taken when raising and lowering indices with the metric.

Before solving the $\omega$ equations, we point out that the trace part of
$\omega_{\mu\,\,b}^{\,a}$ (corresponding to the $U(1)$ part in $U(1,3)$) must
decouple from the other gauge fields. It is thus undetermined and decouples
from the Lagrangian after substituting its equation of motion. It imposes a
condition on the $e_{\mu}^{a}$%
\[
\frac{1}{\sqrt{G}}\partial_{\nu}\left(  \sqrt{G}\left(  e_{a}^{\nu}e^{\mu
a}-e_{a}^{\mu}e^{\nu a}\right)  \right)  \equiv X_{\quad a}^{\mu a}=0.
\]
We can therefore assume, without any loss in generality, that $\omega
_{\mu\,\,b}^{\,a}$ is traceless $\left(  \omega_{\mu\,\,a}^{\,a}=0\right)  .$

The $\omega$-equation gives
\[
\omega_{\kappa\rho}^{\quad\mu}+\omega_{\rho\,\,\kappa}^{\,\,\mu}=\frac{1}%
{8}\delta_{\kappa}^{\mu}\left(  3X_{\,\,\rho\mu}^{\mu}-X_{\,\,\mu\rho}^{\mu
}\right)  +\frac{1}{8}\delta_{\rho}^{\mu}\left(  -X_{\,\,\kappa\mu}^{\mu
}+3X_{\,\,\mu\kappa}^{\mu}\right)  -X_{\,\,\rho\kappa}^{\mu}\equiv
Y_{\,\,\rho\kappa}^{\mu}.
\]
We can rewrite this equation after contracting with $e_{\mu c}e_{\sigma}^{c}$
to get
\[
\omega_{\kappa\rho\sigma}+e_{a}^{\mu}e_{\mu c}e_{\sigma}^{c}\omega
_{\rho\,\,\kappa}^{\,\,\,a}=g_{\sigma\mu}Y_{\,\,\rho\kappa}^{\mu}\equiv
Y_{\sigma\rho\kappa}.
\]
By writing $\omega_{\rho\,\,\kappa}^{\,\,\,a}=\omega_{\rho\nu\kappa}e^{\nu a}$
we get
\[
\left(  \delta_{\kappa}^{\alpha}\delta_{\rho}^{\beta}\delta_{\sigma}^{\gamma
}+g^{\beta\mu}g_{\sigma\mu}\delta_{\rho}^{\alpha}\delta_{\kappa}^{\gamma
}\right)  \omega_{\alpha\beta\gamma}=Y_{\sigma\rho\kappa}.
\]
To solve this equation we have to invert the tensor
\[
M_{\kappa\rho\sigma}^{\alpha\beta\gamma}=\delta_{\kappa}^{\alpha}\delta_{\rho
}^{\beta}\delta_{\sigma}^{\gamma}+g^{\beta\mu}g_{\sigma\mu}\delta_{\rho
}^{\alpha}\delta_{\kappa}^{\gamma}.
\]
In the conventional case when all fields are real, the metric $g_{\mu\nu}$ is
symmetric and $g^{\beta\mu}g_{\sigma\mu}=\delta_{\sigma}^{\beta}$ so that the
inverse of $M_{\kappa\rho\sigma}^{\alpha\beta\gamma}$ is simple. In the
present case, because of the nonsymmetry of $g_{\mu\nu}$ this is fairly
complicated and could only be solved by a perturbative expansion. Writing
$g_{\mu\nu}=G_{\mu\nu}+iB_{\mu\nu}$, and defining $G^{\mu\nu}G_{\nu\rho
}=\delta_{\rho}^{\mu}$ implies that
\begin{align*}
g^{\mu\alpha}g_{\nu\alpha}  &  \equiv\delta_{\nu}^{\mu}+L_{\nu}^{\mu},\\
L_{\nu}^{\mu}  &  =iG^{\mu\rho}B_{\rho\nu}-2G^{\mu\rho}B_{\rho\sigma}%
G^{\sigma\alpha}B_{\alpha\nu}+O(B^{3}).
\end{align*}
The inverse of $M_{\kappa\rho\sigma}^{\alpha\beta\gamma}$ defined by
\[
N_{\alpha\beta\gamma}^{\sigma\rho\kappa}M_{\sigma\rho\kappa}^{\alpha^{\prime
}\beta^{\prime}\gamma^{\prime}}=\delta_{\alpha}^{\alpha^{\prime}}\delta
_{\beta}^{\beta^{\prime}}\delta_{\gamma}^{\gamma^{\prime}}%
\]
is evaluated to give
\begin{align*}
N_{\alpha\beta\gamma}^{\sigma\rho\kappa}  &  =\frac{1}{2}\left(
\delta_{\gamma}^{\sigma}\delta_{\beta}^{\rho}\delta_{\alpha}^{\kappa}%
+\delta_{\beta}^{\sigma}\delta_{\alpha}^{\rho}\delta_{\gamma}^{\kappa}%
-\delta_{\alpha}^{\sigma}\delta_{\gamma}^{\rho}\delta_{\beta}^{\kappa}\right)
\\
&  -\frac{1}{4}\left(  \delta_{\beta}^{\kappa}\delta_{\alpha}^{\sigma
}L_{\gamma}^{\rho}+\delta_{\alpha}^{\kappa}\delta_{\gamma}^{\sigma}L_{\beta
}^{\rho}-\delta_{\gamma}^{\kappa}\delta_{\beta}^{\sigma}L_{\alpha}^{\rho
}\right) \\
&  +\frac{1}{4}\left(  L_{\gamma}^{\kappa}\delta_{\beta}^{\sigma}%
\delta_{\alpha}^{\rho}+L_{\beta}^{\kappa}\delta_{\alpha}^{\sigma}%
\delta_{\gamma}^{\rho}-L_{\alpha}^{\kappa}\delta_{\gamma}^{\sigma}%
\delta_{\beta}^{\rho}\right) \\
&  -\frac{1}{4}\left(  \delta_{\alpha}^{\kappa}L_{\gamma}^{\sigma}%
\delta_{\beta}^{\rho}+\delta_{\gamma}^{\kappa}L_{\beta}^{\sigma}\delta
_{\alpha}^{\rho}-\delta_{\beta}^{\kappa}L_{\alpha}^{\sigma}\delta_{\gamma
}^{\rho}\right)  +O(L^{2}).
\end{align*}
This enables us to write
\[
\omega_{\alpha\beta\gamma}=N_{\alpha\beta\gamma}^{\sigma\rho\kappa}%
Y_{\rho\sigma\kappa}.
\]
It is clear that the leading term reproduces the Einstein-Hilbert action plus
contributions proportional to $B_{\mu\nu}$ and higher order terms. We can
check that in the flat approximation for gravity with $G_{\mu\nu}$ taken to be
$\delta_{\mu\nu}$, the $B_{\mu\nu}$ field gets the correct kinetic terms.
First we write
\[
e_{\mu}^{a}=\delta_{\mu}^{a}-\frac{i}{2}B_{\mu a},\quad e_{\mu a}=\delta_{\mu
}^{a}+\frac{i}{2}B_{\mu a}.
\]
The $\omega_{\mu\,\,a}^{\,\,a}$equation implies the constraint
\[
X_{\,\quad a}^{\mu a}=\partial_{\nu}\left(  e_{a}^{\mu}e^{\nu a}-e_{a}^{\nu
}e^{\mu a}\right)  =0.
\]
This gives the gauge fixing condition $\partial^{\nu}B_{\mu\nu}=0.$ This
equation gave the motivation to Einstein to interpret $B_{\mu\nu}$ as the
electromagnetic field strength. We then evaluate
\[
\omega_{\mu\nu\rho}=-\frac{i}{2}\left(  \partial_{\mu}B_{\nu\rho}%
+\partial_{\nu}B_{\mu\rho}\right)  .
\]
When the $\omega_{\mu\nu\rho}$ is substituted back into the Lagrangian, and
after integration by parts one gets
\[
L=\omega_{\mu\nu\rho}\omega^{\nu\rho\mu}-\omega_{\mu\,\,}^{\,\,\,\mu\rho
}\omega_{\nu\rho}^{\quad\nu}=-\frac{1}{4}B_{\mu\nu}\partial^{2}B^{\mu\nu}.
\]
This is identical to the usual expression $\frac{1}{12}H_{\mu\nu\rho}H^{\mu
\nu\rho},$ where $H_{\mu\nu\rho}=\partial_{\mu}B_{\nu\rho}+\partial_{\nu
}B_{\rho\mu}+\partial_{\rho}B_{\mu\nu}.$ The later developments of
nonsymmetric gravity showed that the occurrence of the trace part of the
spin-connection in a linear form would result in the propagation of ghosts in
the field $B_{\mu\nu}$ \cite{Damour1}. This can be traced to the fact that
there is no gauge symmetry associated with the field $B_{\mu\nu}.$ The
inconsistency is avoided by adding to the action the gauge invariant
cosmological term \cite{Damour2}
\[%
{\displaystyle\int}
d^{4}x\left(  \det e_{\mu a}\det e_{\mu}^{a}\right)  ^{\frac{1}{2}}%
\]
as this provides a mass term to the field $B_{\mu\nu}.$

\section{Unifying space-time and internal symmetries}

{In D-branes, coordinates of space-time become noncommuting and $U(N)$
matrix-valued \cite{Matrix} }%
\[
\left[  X^{i},X^{j}\right]  \neq0.
\]
{ A metric on such spaces will also become matrix-valued. For example in the
case of D-0 branes a matrix model action takes the form \cite{D0}}%
\[
Tr\left(  G_{ij}\left(  X\right)  \partial_{0}X^{i}\partial_{0}X^{j}\right)
.
\]
{At very short distances coordinates of space-time can become noncommuting and
represented by matrices. }

{Developing differential geometry on such spaces is ambiguous. Defining
covariant derivatives, affine connections, contracting indices, will all
depend on the order these operations are performed because of
noncommutativity. Some of these developments lead to inconsistencies such as
the occurrence of higher spin fields \cite{maluf}. In many cases studies were
limited to abelian (commuting) matrices with Fierz-Pauli interactions
\cite{wald}.} More recently the spectral approach was taken by Avramidi
\cite{Avramidi} which implies a well defined order for geometric constructs.
{Experimentally \cite{PDG}, there is only one massless graviton. Therefore in
a consistent $U(N)$ matrix-valued gravity only one massless field should
result with all others corresponding to massive gravitons. The masses of the
gravitons should be acquired through the Higgs mechanism.}

{The lesson we learned in the last section is that one should start with a
large symmetry and break it spontaneously. The minimal non-trivial extension
of $SL(2,\mathbb{C})$ and $U(N)$ is $SL(2N,\mathbb{C}).$ This is a non-compact
group. It can be taken as a gauge group only in the first order formalism, in
analogy with $SL(2,\mathbb{C}).$}{ The vierbein $e_{\mu}^{a}$ and the
spin-connection $\omega_{\mu}^{\;ab}$ are conjugate variables related by the
zero torsion condition. The number of conditions in $T_{\mu\nu}^{\;\;a}=0$ is
equal to the number of independent components of $\omega_{\mu}^{\;ab}$, which
can be determined completely in terms of $e_{\mu}^{a}.$ }The $SL(2N,\mathbb{C}%
)$ gauge field can be expanded in the Dirac basis in the form \cite{ISS}%
,\cite{colored}%
\[
A_{\mu}=ia_{\mu}+\gamma_{5}b_{\mu}+\frac{i}{4}\omega_{\mu}^{\;ab}\sigma_{ab},
\]
where {%
\begin{align*}
a_{\mu}  &  =a_{\mu}^{I}\lambda^{I},\hspace{0.3in}b_{\mu}=b_{\mu}^{I}%
\lambda^{I},\hspace{0.3in}I=1,\cdots,N^{2}-1,\\
\omega_{\mu}^{\;ab}  &  =\omega_{\mu}^{\;abi}\lambda^{i},\hspace{0.3in}i=0,I.
\end{align*}
and }$\lambda^{i}$ are the $U(N)$ Gell-Mann matrices. The analogue of $e_{\mu
}^{a}\gamma_{a}$ is {%
\[
L_{\mu}=e_{\mu}^{a}\gamma_{a}+f_{\mu}^{a}\gamma_{5}\gamma_{a},
\]
}where $e_{\mu}^{a}$ and $f_{\mu}^{a}$ are $U(N)$ matrices. This is equivalent
to having complex matrix gravity. The zero torsion condition {%
\[
T=dL+LA+AL=0,
\]
}will give two sets of conditions {%
\[
T_{\mu\nu}^{\;\;a}=0,\hspace{0.3in}T_{\mu\nu}^{\;\;a5}=0,
\]
}which will overdetermined the variables $\omega_{\mu}^{\;ab}.$

{The correct approach \cite{colored} is to consider $SL(2N,\mathbb{C}%
)\mathbb{\times}SL(2N,\mathbb{C}),$ or equivalently the complex extension of
$SL(2N,\mathbb{C)}$ as was done by Isham, Salam and Strathdee \cite{ISS} for
the massive spin-2 nonets. In this case{%
\[
a_{\mu}=a_{\mu}^{1}+ia_{\mu}^{2},\hspace{0.3in}b_{\mu}=b_{\mu}^{1}+ib_{\mu
}^{2},\hspace{0.3in}\omega_{\mu}^{\;ab}=B_{\mu}^{\;ab}+iC_{\mu}^{\;ab},
\]
}and the torsion zero constraints are enough to determine $B_{\mu}^{ab}$ and
$C_{\mu}^{ab}$ in terms of $e_{\mu}^{a}$, $f_{\mu}^{a},$ $a_{\mu}$ and
$b_{\mu}.$ One can write, almost uniquely, a metric independent gauge
invariant action which will correspond to massless $U(N)$ gravitons{%
\[
\int\limits_{M}Tr\left(  i\left(  \alpha+\beta\gamma_{5}\right)  LL^{^{\prime
}}F+i\left(  \overline{\alpha}+\overline{\beta}\gamma_{5}\right)  L^{^{\prime
}}L\overline{F}+\left(  i\lambda+\gamma_{5}\eta\right)  LL^{^{\prime}%
}LL^{^{\prime}}\right)  ,
\]
}where $L^{^{\prime}}$ is related to $L.$ For illustration, the form of this
action in the }$N=1$ \ case is
\begin{align*}
&  -\frac{1}{2}\int\limits_{M}d^{4}x\epsilon^{\mu\nu\kappa\lambda}\left(
\left(  \left(  \alpha_{2}-\beta_{1}\right)  e_{\mu a}e_{\nu b}+\frac{1}%
{2}\left(  \alpha_{1}+\beta_{2}\right)  \epsilon_{abcd}e_{\mu}^{c}e_{\nu}%
^{d}\right)  B_{\kappa\lambda}^{\;\;\;ab}\right.  \\
&  \qquad\qquad+\left(  \left(  \alpha_{2}+\beta_{1}\right)  f_{\mu a}f_{\nu
b}-\frac{1}{2}\left(  \alpha_{1}-\beta_{2}\right)  \epsilon_{abcd}f_{\mu}%
^{c}f_{\nu}^{d}\right)  C_{\kappa\lambda}^{\;\;ab}\\
&  \qquad\qquad\left.  +\epsilon_{abcd}\left(  \left(  \lambda-\eta\right)
e_{\mu}^{a}e_{\nu}^{b}e_{\kappa}^{c}e_{\lambda}^{d}+\left(  \lambda
+\eta\right)  f_{\mu}^{a}f_{\nu}^{b}f_{\kappa}^{c}f_{\lambda}^{d}\right)
\right)  .
\end{align*}
where $B_{\kappa\lambda}^{\;\;ab}$ $C_{\kappa\lambda}^{\;ab}$ are the
curvatures associated with $B_{\mu}^{\;ab}$ and $C_{\mu}^{\;ab}$ respectively.
{To give masses to the spin-2 fields, introduce the Higgs fields $H$ and
$H^{^{\prime}}$ transforming as $L$ and $L^{^{\prime}}$ and constrained in
such a way as to break the symmetry non-linearly from $SL(2N,\mathbb{C)\times
}SL(2N,\mathbb{C)}$ to $SL(2,\mathbb{C)}$} \cite{CWZ}.{ We can add the mass
terms{%
\[
\int\limits_{M}Tr\left(  \left(  i\tau+\gamma_{5}\xi\right)  LH^{^{\prime}%
}LH^{^{\prime}}LL^{^{\prime}}+\left(  i\rho+\gamma_{5}\xi\right)
HL^{^{\prime}}HL^{^{\prime}}LL^{^{\prime}}\right)  .
\]
}Some of the relevant terms in the quadratic parts of the action are, in
component form \cite{colored},}{%
\begin{align*}
&  \int d^{4}x\epsilon^{\mu\nu\kappa\lambda}Tr\left(  \alpha_{1}\left\{
E_{\mu}^{a},E_{\nu a}^{^{\prime}}\right\}  a_{\kappa\lambda}^{2}+\alpha
_{2}\left\{  F_{\mu}^{a},F_{\nu a}^{^{\prime}}\right\}  b_{\kappa\lambda}%
^{2}\right)  \\
&  \hspace{0.3in}\hspace{0.3in}+\epsilon_{abcd}Tr\left(  \beta_{1}\left\{
E_{\mu}^{a},E_{\nu}^{^{\prime}b}\right\}  B_{\kappa\lambda}^{cd}+\beta
_{2}\left\{  F_{\mu}^{a},F_{\nu}^{^{\prime}b}\right\}  C_{\kappa\lambda}%
^{cd}+\right.  \\
&  \left.  +\gamma_{1}E_{\mu}^{a}E_{\nu}^{^{\prime}b}E_{\kappa}^{c}E_{\lambda
}^{^{\prime}d}+\gamma_{2}F_{\mu}^{a}F_{\nu}^{^{\prime}b}F_{\kappa}%
^{c}F_{\lambda}^{^{\prime}d}+\delta_{1}E_{\mu}^{a}E_{\nu}^{^{\prime}%
b}E_{\kappa}^{c}F_{\lambda}^{d}+\delta_{2}F_{\mu}^{a}F_{\nu}^{^{\prime}%
b}F_{\kappa}^{c}E_{\lambda}^{d}\right)  .
\end{align*}
This action is complicated because all expressions are matrix valued.
Equations are solved perturbatively. The action can be determined to second
order in the fields, and the spectrum found to be given by two sets of $SU(N)$
matrix-valued massive gravitons, plus two singlets of gravitons, one massless
and the other is massive, as well as $SU(N)\times SU(N)$ gauge fields.}

{We decompose $E_{\mu a}^{I}$ into symmetric and antisymmetric parts}{%
\[
E_{\mu a}^{I}=S_{\mu a}^{I}+T_{\mu a}^{I},
\]
}{where $S_{\mu a}^{I}=S_{a\mu}^{I}$ is symmetric and $T_{\mu a}^{I}=-T_{a\mu
}^{I}$ is antisymmetric. The symmetric part propagates while the antisymmetric
part $T_{\mu\nu}^{I}$ couples to the Yang-Mills fields and act as auxiliary
fields to give them kinetic energies. For example, besides the quadratic terms
for $T^{\mu\nu I}$ coming from the mass terms, we have} {%
\[
\int\limits_{M}d^{4}x\left(  \partial_{\mu}a_{\nu}^{1I}-\partial_{\nu}a_{\mu
}^{1I}\right)  T^{\mu\nu I},
\]
}{as well as similar couplings to $a_{\mu}^{2I},\ b_{\mu}^{1I},\ b_{\mu}^{2I}%
$. }By eliminating the field $T_{\mu\nu}^{I}$ the fields {$a_{\mu}%
^{1I},\ a_{\mu}^{2I},\ b_{\mu}^{1I},\ b_{\mu}^{2I}$} would acquire the regular
$SU(N)$ Yang-Mills gauge field strengths. {A detalied study of this system is
carried in \cite{colored}. \ Again, this shows the effectiveness of the gauge
principle in generalizing unambiguosly the metric and gravitational
interactions to become matrix valued. This is to be contrasted with the
geometric approach, which is plagued with ambiguities.}

\section{Supergravity from gauging graded Lie algebras}

The gauge formulation of gravity is simpler than the geometrical formulation.
However, the simplifications are much more apparent in the derivation of the
supergravity action based on gauging the supersymmetry algebra. In 1976 the
supergravity action was constructed by extending the local supersymmetric
invariance to the Einstein action by using the Noether's method \cite{Ferrara}%
. This is a perturbative approach of insuring the invariance order by order,
up to quartic fermionic terms and is very complicated. It was natural to
attempt construct this theory using the gauge approach, and indeed this was
done soon after and resulted in the most elegant formulation of supergravity.
The starting point is to consider the supersymmetry algebra. This is the
graded extension of the Poincar\'{e} algebra by adding the following
commutation relations \cite{CW}%
\begin{align*}
\left[  S_{\alpha},J_{ab}\right]   &  =\frac{i}{2}\left(  \gamma_{ab}\right)
_{\alpha}^{\beta}S_{\beta},\\
\left[  S_{\alpha},P_{a}\right]   &  =0,\\
\left\{  S_{\alpha},S_{\beta}\right\}   &  =-\left(  \gamma^{a}C\right)
_{\alpha\beta}P_{a},
\end{align*}
where $P_{a},$ $J_{ab}$ and $S_{\alpha}$ are, respectively, the translation,
rotation and fermionic generators and $C_{\alpha\beta}$ is the charge
conjugation matrix. Demanding invariance under local supersymmetry
transformations, requires introducing the covariant derivative%
\[
D_{\mu}=\partial_{\mu}+e_{\mu}^{a}P_{a}+\omega_{\mu}^{\;ab}J_{ab}%
+\overline{\psi}_{\mu}^{\alpha}S_{\alpha}%
\]
where $e_{\mu}^{a},$ $\omega_{\mu}^{\;ab}$ and $\overline{\psi}_{\mu}^{\alpha
}$ are now gauge fields with the following fermionic transformations%
\begin{align*}
\delta e_{\mu}^{a} &  =\overline{\eta}\gamma^{a}\psi_{\mu}\\
\delta\omega_{\mu}^{\;ab} &  =0,\\
\delta\overline{\psi}_{\mu}^{\alpha} &  =\left[  \overline{\eta}\left(
\overleftarrow{\partial}_{\mu}-\frac{1}{4}\gamma_{ab}\omega_{\mu}%
^{\;ab}\right)  \right]  ^{\alpha}.
\end{align*}
The fermionic gauge parameters $\eta$ are dependent on the coordinates
$x^{\mu}.$ The field strengths are found by computing the commutator%
\[
\left[  D_{\mu},D_{\nu}\right]  =C_{\mu\nu}^{\;\;\;a}P_{a}+\frac{1}{4}%
R_{\mu\nu}^{\;\;\;ab}\gamma_{ab}+\overline{D}_{\mu\nu}^{\alpha}S_{\alpha},
\]
where
\begin{align*}
C_{\mu\nu}^{\;\;\;a} &  =\partial_{\mu}e_{\nu}^{a}+\omega_{\mu}^{\;ab}e_{\nu
b}-\frac{1}{2}\overline{\psi}_{\mu}\gamma^{a}\psi_{\nu}-\mu\leftrightarrow
\nu,\\
R_{\mu\nu}^{\quad ab} &  =\partial_{\mu}\omega_{\nu}^{\;ab}+\omega_{\mu
}^{\;ac}\omega_{\mu c}^{\quad b}-\mu\leftrightarrow\nu,\\
\overline{D}_{\mu\nu}^{\alpha} &  =\overline{\psi}_{\left[  \nu\right.
}\left(  \overleftarrow{\partial}_{\left.  \mu\right]  }-\frac{1}{4}%
\gamma_{ab}\omega_{\left.  \mu\right]  }^{\;ab}\right)  ^{\alpha}%
-\mu\leftrightarrow\nu.
\end{align*}
The field strengths transform covariantly under supersymmetry transformations%
\begin{align*}
\delta C_{\mu\nu}^{\quad a} &  =\overline{\eta}\gamma^{a}\psi_{\mu\nu},\\
\delta R_{\mu\nu}^{\quad ab} &  =0,\\
\delta\overline{\psi}_{\mu\nu}^{\alpha} &  =\frac{1}{4}\left(  \overline{\eta
}\gamma_{ab}R_{\mu\nu}^{\;\;\;ab}\right)  ^{\alpha}%
\end{align*}
In analogy with the bosonic case, we assume that the generalized torsion
$C_{\mu\nu}^{\;\;a}$, which is the field strength associated with
translational symmetry, to vanish. It is then necessary to modify the
transformations of $\omega_{\mu}^{\;ab}$ in order to maintain this condition,
thus we must take
\[
\delta^{^{\prime}}\omega_{\mu ab}=-\frac{1}{4}\left(  \overline{\eta}%
\gamma_{a}D_{\mu b}-\overline{\eta}\gamma_{b}D_{\mu a}+e_{\mu}^{c}e_{a}^{\rho
}e_{b}^{\sigma}\overline{\eta}\gamma_{c}D_{\rho\sigma}\right)  .
\]
The action is then constructed as function of the remaining field strengths:%
\[
I=\frac{1}{4}%
{\displaystyle\int}
d^{4}x\epsilon^{\mu\nu\rho\sigma}\left(  \epsilon_{abcd}e_{\mu}^{a}e_{\nu}%
^{b}R_{\rho\sigma}^{\;\;\;cd}+i\alpha e_{\mu}^{a}\overline{\psi}_{\nu}%
\gamma_{5}\gamma_{a}D_{\rho\sigma}\right)
\]
where the parameter $\alpha$ is determined by the requirement that the
$\omega_{\mu}^{\;ab}$ equation of motion, which appears linearly and
quadratically, should result in the equation $C_{\mu\nu}^{\;\;a}=0.$ This
fixes the parameter $\alpha=1.$ Using the observation that the variation of
the action with respect to $\omega_{\mu}^{\;ab}$ drops out because
\[
\left.  \frac{\delta I}{\delta\omega_{\mu}^{\;ab}}\right\vert _{e,\psi}=0,
\]
proving the invariance of the action under supersymmetry transformations
become an easy matter. Varying the Einstein term only gives one term
\[
2\epsilon_{abcd}\overline{\eta}\gamma^{a}\psi_{\mu}e_{\nu}^{b}R_{\rho\sigma
}^{\;\;\;cd},
\]
while varying the fermionic term results in three terms
\[
i\overline{\eta}\gamma^{a}\psi_{\mu}\overline{\psi}_{\nu}\gamma_{5}\gamma
_{a}D_{\rho\sigma}+ie_{\mu}^{a}\overline{\eta}\overleftarrow{D}_{\nu}%
\gamma_{5}\gamma_{a}D_{\rho\sigma}-\frac{i}{4}e_{\mu}^{a}\overline{\psi}_{\nu
}\gamma_{5}\gamma_{a}\gamma_{cd}\eta R_{\rho\sigma}^{\;\;\;cd}.
\]
Integrating by parts the middle term gives two terms, one proportional to
$D_{\left[  \mu\right.  }e_{\left.  \nu\right]  }^{a}$ which can be equated to
$\frac{1}{2}\overline{\psi}_{\mu}\gamma^{a}\psi_{\nu}$ using the generalized
torsion condition, while the other is proportional to $D_{\left[  \nu\right.
}D_{\left.  \rho\sigma\right]  }$ which by the Bianchi identity is equal to
$\frac{1}{4}R_{\left[  \nu\rho\right.  }^{\quad\;\;cd}\gamma_{cd}\psi_{\left.
\sigma\right]  }.$ One can show by a simple Fierz reshuffle that these two
terms cancel the first and third terms in the above expression. The simplicity
of this derivation \cite{CW} should be contrasted with the extreme complexity
of the Noether's method which was originally used to derive this Lagrangian
\cite{Ferrara}.

Another method of deriving the supergravity action is based on the observation
that the supersymmetry algebra could be obtained by an Inon\"{u}-Wigner
contraction of the Orthosymplectic algebra $OSP(4,1)$ \cite{mcdowel}%
,\cite{ortho},\cite{ch78}. The group is defined as the set of linear
transformations which leave invariant the bilinear form
\[
\left(  z,z\right)  =z^{A}\eta_{AB}z^{B},
\]
where the linear space $\left\{  z^{A}=\theta^{\alpha},z\right\}  $ comprise
one commuting coordinate $z^{5}=z$ and four anticommuting coordinates
$z^{A}=\theta^{\alpha},$ $A=1,\cdots,4$. The matrix $\eta_{AB}$ is given by
\[
\eta_{AB}=\left(
\begin{array}
[c]{cc}%
C_{\alpha\beta} & 0\\
0 & 1
\end{array}
\right)  ,
\]
where $C_{\alpha\beta}$ is an antisymmetric root of the unit matrix
\cite{ortho}. Indices are raised and lowered with $\eta_{AB}$ and its inverse
$\eta^{AB},$ where $\eta^{AB}\eta_{BC}=\delta_{C}^{A},$ thus
\[
z_{A}=\eta_{AB}z^{B},\quad z^{A}=\eta^{AB}z_{B}.
\]
The tensor $\phi_{A}^{\;B},$ a representation of $OSP(1,4)$, transforms like
$z_{A}z^{B}$, contains three distinct irreducible representations. These are
the graded trace $(-1)^{a}\phi_{A}^{\;A},$ the graded symmetric and traceless
part, and the graded antisymmetric part. The grading number $a$ for $z^{A}$ is
defined by
\begin{align*}
a  &  =0,\quad A=5\\
a  &  =1,\quad A=1,\cdots,4
\end{align*}
so that $z^{A}z^{B}=\left(  -1\right)  ^{ab}z^{B}z^{A}$ \ and the graded
symmetric and antisymmetric tensors are defined by
\begin{align*}
\phi_{AB}^{\left(  s\right)  }  &  =\left(  -1\right)  ^{ab}\phi_{BA}^{\left(
s\right)  },\\
\phi_{AB}^{\left(  a\right)  }  &  =-\left(  -1\right)  ^{ab}\phi
_{BA}^{\left(  a\right)  },\\
(-1)^{a}\phi_{A}^{\left(  s\right)  \;A}  &  =0,
\end{align*}
where $\phi_{AB}=\phi_{A}^{\;C}\eta_{CB}$.

The matrix decomposition of these representations are given by \cite{ortho}%
\begin{align*}
\phi_{A}^{(s)B} &  =\left(
\begin{array}
[c]{cc}%
\left(  \frac{1}{4}\varphi+\gamma\pi+\gamma_{a}\gamma_{5}v^{a}\right)
_{\alpha}^{\beta} & \lambda_{\alpha}\\
-\overline{\lambda}^{\alpha} & \phi
\end{array}
\right)  ,\\
\phi_{A}^{(a)B} &  =\left(
\begin{array}
[c]{cc}%
\left(  \gamma_{a}\phi^{a}+\frac{1}{4}\gamma_{ab}\phi^{ab}\right)  _{\alpha
}^{\beta} & \lambda_{\alpha}\\
\overline{\lambda}^{\alpha} & 0
\end{array}
\right)  ,
\end{align*}
where the spinors used are Majorana, $\lambda_{\alpha}=C_{\alpha\beta
}\overline{\lambda}^{\beta}.$ The antisymmetric representation is the adjoint
representation. Demanding local $OSP(1,4)$ gauge invariance is done along
similar lines to the pure bosonic case. First introduce gauge potentials in
the adjoint representation
\[
\Phi_{\mu}=\left(
\begin{array}
[c]{cc}%
\left(  \kappa^{-1}\left(  i\gamma_{a}\right)  _{\alpha}^{\beta}e_{\mu}%
^{a}+\frac{1}{4}\left(  \gamma_{ab}\right)  _{\alpha}^{\beta}\omega_{\mu
}^{\;ab}\right)   & \kappa^{-\frac{1}{2}}\psi_{\mu\alpha}\\
\kappa^{-\frac{1}{2}}\overline{\psi}_{\mu}^{\alpha} & 0
\end{array}
\right)
\]
which transforms according to
\[
\Phi_{\mu}=\Omega\Phi_{\mu}\Omega^{-1}+\Omega\partial_{\mu}\Omega^{-1}%
\]
where $\Omega$ are gauge parameters in the adjoint representation. The field
strengths of the gauge potentials $\Phi_{\mu}$ are defined by%
\begin{align*}
\Phi_{\mu\nu} &  =\partial_{\mu}\Phi_{\nu}-\partial_{\nu}\Phi_{\mu}+\left[
\Phi_{\mu},\Phi_{\nu}\right]  \\
&  =\left(
\begin{array}
[c]{cc}%
\left(  \kappa^{-1}\left(  i\gamma_{a}\right)  _{\alpha}^{\beta}C_{\mu\nu
}^{\;\;a}+\frac{1}{4}\left(  \gamma_{ab}\right)  _{\alpha}^{\beta}F_{\mu\nu
}^{\;\;ab}\right)   & \kappa^{-\frac{1}{2}}\chi_{\mu\nu\alpha}\\
\kappa^{-\frac{1}{2}}\overline{\chi}_{\mu\nu}^{\alpha} & 0
\end{array}
\right)
\end{align*}
where
\begin{align*}
C_{\mu\nu}^{\;\;\;a} &  =\partial_{\mu}e_{\nu}^{a}+\omega_{\mu}^{\;ab}e_{\nu
b}-\frac{i}{4}\overline{\psi}_{\mu}\gamma^{a}\psi_{\nu}-\mu\leftrightarrow
\nu,\\
F_{\mu\nu}^{\quad\;ab} &  =R_{\mu\nu}^{\quad\;ab}-4\kappa^{-2}\left(  e_{\mu
}^{a}e_{\nu}^{b}-e_{\mu}^{a}e_{\nu}^{b}\right)  -\kappa^{-1}\overline{\psi
}_{\mu}\gamma^{ab}\psi_{\nu},\\
\chi_{\mu\nu}^{\alpha} &  =D_{\mu\nu}^{\alpha}+2\kappa^{-1}\left(  i\gamma
_{a}\right)  _{\beta}^{\alpha}e_{\left[  \mu\right.  }^{a}\psi_{\left.
\nu\right]  b}^{\beta}.
\end{align*}
The orthosymplectic algebra $OSP(4,1)$ contracts to the supersymmetry algebra
by taking the limit $\kappa\rightarrow\infty.$ This corresponds to the
contraction of a de Sitter space with infinite radius to give Minkowski space.
The component gauge fields transform according to
\begin{align*}
\delta e_{\mu}^{a} &  =\partial_{\mu}\varepsilon^{a}+\omega_{\mu}%
^{\;ab}\varepsilon_{b}+\overline{\eta}\gamma^{a}\psi_{\mu}+\omega^{ab}e_{\mu
b}\\
\delta\omega_{\mu}^{\;ab} &  =\partial_{\mu}\omega^{ab}+2\omega^{\left[
a\right.  c}\omega_{\mu c}^{\quad\left.  b\right]  }+2\kappa^{-2}e_{\mu
}^{\left[  a\right.  }\varepsilon^{\left.  b\right]  }+2\kappa^{-1}%
\overline{\eta}\gamma^{ab}\psi_{\mu},\\
\delta\psi_{\mu} &  =\partial_{\mu}\eta+\frac{1}{4}\omega_{\mu}^{\;ab}%
\gamma_{ab}\eta+\frac{1}{4}\omega^{\;ab}\gamma_{ab}\psi_{\mu}+i\kappa
^{-1}e_{\mu}^{a}\gamma_{a}\eta+i\kappa^{-1}\varepsilon^{a}\gamma_{a}\psi_{\mu}%
\end{align*}
Imposing the constraint on the field strength of the translation generators
\[
C_{\mu\nu}^{\;\;\;a}=0
\]
requires the modification of the transformations of $\omega_{\mu}^{\;ab}$ in
order to maintain this constraint. The invariant action is then formed from
the other non-vanishing components of $\Phi_{\mu\nu}$, and is given by%
\[
I=%
{\displaystyle\int}
d^{4}x\epsilon^{\mu\nu\rho\sigma}\left(  \epsilon_{abcd}F_{\mu\nu}^{\quad
\;ab}F_{\rho\sigma}^{\quad\;cd}+\alpha\overline{\chi}_{\mu\nu}\gamma_{5}%
\chi_{\rho\sigma}\right)  ,
\]
where the parameter $\alpha$ is fixed by gauge invariance to be
\[
\alpha=\frac{i}{2}.
\]
The above action splits into five parts ranging in powers of $\kappa$ from
$\kappa^{0}$ to $\kappa^{-4}.$ The first is the Gauss-Bonnet topological
invariant%
\[
I^{\left(  0\right)  }=\frac{1}{4}%
{\displaystyle\int}
d^{4}x\epsilon^{\mu\nu\rho\sigma}\epsilon_{abcd}R_{\mu\nu}^{\;\;\;ab}%
R_{\rho\sigma}^{\;\;\;cd}.
\]
The second part is a boundary term and is given by%
\[
I^{\left(  -1\right)  }=i\kappa^{-1}%
{\displaystyle\int}
d^{4}x\epsilon^{\mu\nu\rho\sigma}\overline{D}_{\mu\nu}\gamma_{5}D_{\rho\sigma
}.
\]
The third part is the supergravity action
\[
I^{\left(  -2\right)  }=4\kappa^{-2}%
{\displaystyle\int}
d^{4}x\epsilon^{\mu\nu\rho\sigma}\left(  \epsilon_{abcd}e_{\mu}^{a}e_{\nu}%
^{b}R_{\rho\sigma}^{\;\;\;cd}+ie_{\mu}^{a}\overline{\psi}_{\nu}\gamma
_{5}D_{\rho\sigma}\right)  .
\]
The fourth part part is a mass-like term for the gravitino
\[
I^{\left(  -2\right)  }=4i\kappa^{-3}%
{\displaystyle\int}
d^{4}x\epsilon^{\mu\nu\rho\sigma}\epsilon_{abcd}e_{\mu}^{a}e_{\nu}%
^{b}\overline{\psi}_{\rho}\gamma_{5}\psi_{\sigma},
\]
and the last part is a cosmological term
\[
I^{\left(  -2\right)  }=16\kappa^{-4}%
{\displaystyle\int}
d^{4}x\epsilon^{\mu\nu\rho\sigma}\epsilon_{abcd}e_{\mu}^{a}e_{\nu}^{b}e_{\rho
}^{c}e_{\sigma}^{d}.
\]
The total Lagrangian is that of supergravity with a cosmological constant,
mass like term for the gravitino and boundary terms. Using the fact that the
first two terms are boundary terms and can be discarded, we notice that in the
limit $\kappa\rightarrow\infty,$ the action reduces to the supergravity action
$I^{\left(  -2\right)  }$ which is invariant under the supersymmetry algebra.
This is to be expected because the supersymmetry algebra is obtained from the
orthosymplectic algebra $OSP(4,1)$ by an Inon\"{u}-Wigner contraction by
taking the limit $\kappa\rightarrow\infty.$ This analysis shows the power of
the gauge idea, and reduce an intractable calculation to the evaluation of a
simple trace. This was just the starting point for gauging graded Lie groups,
and many applications of this idea followed soon after. This idea played an
important role in the construction of extended supergravities \cite{cremmer}
and conformal supergravity \cite{Townsend}.

\section{Topological gravity in odd dimensions}

It was shown by Witten \cite{Witten3d} that the Einstein-Hilbert action with
or without a cosmological term for three dimensional gravity can be derived in
the first order formalism as a gauge theory of $SO(1,4),$ $SO(2,3),$ or
$ISO(1,3)$. The action is of the Chern-Simons type and is renormalizable. This
formalism could be generalized to all odd dimensions. Consider the $2n+1$
Chern-Simons form $\omega_{2n+1}$ defined by \cite{Zumino}%
\[
\omega_{2n+1}=(n+1)%
{\displaystyle\int\limits_{0}^{1}}
\delta t\left\langle A\left(  tdA+t^{2}A^{2}\right)  ^{n}\right\rangle ,
\]
where $A$ is the gauge field for one of the gauge groups $ISO(1,2n),$
$SO(1,2n+1)$ or $SO(2,2n)$ depending on whether we want to gauge the
Poincar\'{e} de Sitter or anti de Sitter groups \cite{topo},\cite{topo2} in
$2n+1$ dimensions. In the definition of the bracket $\left\langle
\cdots\right\rangle $ it is essential to use the $(n+1)$ group-invariant form
\cite{Witten3d}%
\[
\left\langle J_{A_{1}B_{1}}J_{A_{2}B_{2}}\cdots J_{A_{n}B_{n}}\right\rangle
=\epsilon_{A_{1}B_{1}A_{2}B_{2}\cdots A_{n}B_{n}},
\]
where $J_{AB}$ is the group generator and $A,B=0,1,\cdots,2n+1.$ The action is
taken to be
\[
I_{2n+1}=k%
{\displaystyle\int\limits_{M^{2n+1}}}
\omega_{2n+1}.
\]
Under a gauge transformation the gauge field $A$ transforms according to
\[
A^{g}=g^{-1}Ag+g^{-1}dg
\]
which implies that the Chern-Simons form transforms to
\[
\omega_{2n+1}^{g}=\omega_{2n+1}+d\alpha_{2n}+\left(  -1\right)  ^{n}%
\frac{\left(  n\right)  !\left(  n+1\right)  !}{\left(  2n+1\right)
!}\left\langle \left(  g^{-1}dg\right)  ^{2n+1}\right\rangle ,
\]
where $\alpha_{2n}$ is a two form which is a function of $A$ and $g^{-1}dg,$
and the last term is proportional to the winding number. For the groups
$ISO(1,2n),$ $SO(1,2n+1)$ or $SO(2,2n)$ the winding number is proportional to
torsion, and thus vanishes. For example in the case of five-dimensional
Chern-Simons form the homotopy elements are
\begin{align*}
\pi_{5}\left(  SO(1,5)\right)   &  =\pi_{5}\left(  SO(5)\right)  =Z_{2},\\
\pi_{5}\left(  SO(2,4)\right)   &  =\pi_{5}\left(  SO(4)\right)  =Z_{2}+Z_{2}.
\end{align*}
Thus for manifolds without boundary the Chern-Simons action is gauge
invariant, which implies that the constant $k$ is not quantized, which is
desirable if the theory is to describe gravity. For manifolds with boundary
the action is invariant provided that $A$ or $g^{-1}dg$ vanish at the
boundary. We shall assume that the manifold $M_{2n+1\text{ }}$is without boundary.

To make the connection to gravity we identify
\[
A^{ab}=\omega^{ab},\quad A^{a,2n+1}=e^{a},\quad a=0,1,\cdots,2n.
\]
The remarkable thing is that the above Chern-Simons action when expressed in
terms of the fields $e^{a}$ and $\omega^{ab}$ takes the form%
\begin{align*}
I_{2n+1}  &  =%
{\displaystyle\int\limits_{M_{2n+1}}}
{\displaystyle\sum\limits_{l=0}^{n}}
\frac{1}{2l+1}\lambda^{l}\left(
\begin{array}
[c]{c}%
n\\
l
\end{array}
\right)  \cdot\\
&  \qquad\epsilon_{a_{1}a_{2}\cdots a_{2n+1}}R^{a_{1}a_{2}}\wedge\cdots
R^{a_{2n-2l-1}a_{2n-2l}}\wedge e^{a_{2n-2l+1}}\wedge\cdots\wedge e^{a_{2n+1}},
\end{align*}
where
\[
\lambda=\left\{
\begin{array}
[c]{c}%
1\quad\text{for }SO(2,2n)\\
-1\quad\text{for }SO(1,2n+1)\\
0\quad\text{for }ISO(1,2n)
\end{array}
\right.  .
\]
Thus the Chern-Simons action in odd dimensions is seen to be the sum of Euler
densities with fixed coefficients \cite{topology}. Notice that in the
$ISO(1,2n)$ case only one term in the action remains
\[
I_{2n+1}=%
{\displaystyle\int\limits_{M_{2n+1}}}
\epsilon_{a_{1}a_{2}\cdots a_{2n+1}}R^{a_{1}a_{2}}\wedge\cdots R^{a_{2n-1}%
a_{2n}}\wedge e^{a_{2n+1}}.
\]
The variational equations of $A^{aB}$ have a simple structure%
\[
\epsilon_{A_{1}B_{1}\cdots A_{n}B_{n}}F^{A_{1}B_{1}}\wedge\cdots\wedge
F^{A_{n}B_{n}}=0
\]
where
\[
F^{AB}=dA^{AB}+A^{AC}\wedge A_{C}^{\quad B}.
\]
Decomposing this equation with respect to the new variables gives two
equations%
\begin{align*}
\epsilon_{a_{1}a_{2}\cdots a_{2n+1}}\left(  R^{a_{1}a_{2}}+\lambda e^{a_{1}%
}e^{a_{2}}\right)  \wedge\cdots\wedge\left(  R^{a_{2n-1}a_{2n}}+\lambda
e^{a_{2n-1}}e^{a_{2n}}\right)   &  =0,\\
\epsilon_{a_{1}a_{2}\cdots a_{2n+1}}T^{a_{1}}\wedge\left(  R^{a_{2}a_{3}%
}+\lambda e^{a_{2}}e^{a_{3}}\right)  \wedge\cdots\wedge\left(  R^{a_{2n-2}%
a_{2n-1}}+\lambda e^{a_{2n-2}}e^{a_{2n-1}}\right)   &  =0,
\end{align*}
where $T^{a}$ is the torsion given by
\[
T^{a}=de^{a}+\omega^{ab}\wedge e_{b}.
\]
This scheme could be generalized to include fermions by gauging the graded
extensions of the Poincar\'{e} and de-Sitter groups. The supersymmetric
extension of the Poincar\'{e} group is known to exist in all dimensions, but
those extending the de Sitter groups are limited. The list stops at the
supersymmetric extension of $O(2,6)$ corresponding to a seven-dimensional
space-time in this framework. The supergroups relevant to space-time
dimensions of four or more are \cite{Nahm}
\[
\left\{
\begin{array}
[c]{c}%
\left(  O(2,3)\oplus O(N),(4,N)\right)  ,\quad N=1,2,\cdots\\
\left(  O(1,4)\oplus U(1),\left(  4+\overline{4}\right)  \right)
\end{array}
\right\}  \qquad D=4
\]%
\[
\left\{
\begin{array}
[c]{c}%
\left(  O(2,3)\oplus O(N),\left(  4,N\right)  +\left(  \overline{4}%
,\overline{N}\right)  \right)  ,\quad N=1,2,\cdots\\
\left(  O(2,4)\oplus SU(4),\left(  4+4\right)  +(\overline{4},\overline
{4})\right)
\end{array}
\right\}  \qquad D=5
\]%
\[
\left\{
\begin{array}
[c]{c}%
\left(  O(1,6)\oplus SU(2),(8,2)\right) \\
\left(  O(2,5)\oplus SU(2),\left(  8,2\right)  \right)
\end{array}
\right\}  \qquad D=6
\]%
\[
\left(  O(2,6)\oplus SU(N,q),\left(  8,2N\right)  \right)  ,\quad
N=1,2,\cdots\qquad D=7
\]
In this notation, the first part in the parentheses gives the bosonic parts of
the group while the second part specifies the fermionic representations with
respect to these bosonic groups. The supergravity actions are constructed
using the Chern-Simons forms in odd dimensions with the gauge fields taking
values in the graded Lie algebras corresponding to that dimension. The graded
group invariant is now constructed using the supertrace. The
super-Poincar\'{e} groups are easier to deal with and a general formula can be
given. In dimensions higher than seven, the smallest extensions of the de
Sitter groups are the orthosymplectic groups, which have many more generators
than the minimum required. There are many further developments in this
direction \cite{Banados}, for a review see \cite{zanelli}.We therefore see
that the gauge principle is very easy and powerful in determining topological
theories of gravity in odd dimension. This is to be contrasted with the
geometrical approach where such constructions are quite difficult.

\section{Noncommutative gravity}

Open string theories as well as D-branes in the presence of a background
antisymmetric $B$-field give rise to noncommutative effective field theories
\cite{Matrix},\cite{SW}. This is equivalent to field theories deformed with
the star product \cite{AFS}. The primary example of this is noncommutative
$U(N)$ Yang-Mills theory. It is natural then to ask whether it is possible to
deform Einstein's gravity with the star product. This is not easy to do in a
geometrical setting, although there has been some recent progress
\cite{wess2}. On the other hand this is possible in the gauge approach using
the same methods used before. To construct a noncommutative gravitational
action in four dimensions one proceeds as follows. First the gauge field
strength of the noncommutative gauge group $SO(4,1)$ is taken. This is
followed by an Inon\"{u}-Wigner contraction to the group $ISO(3,1)$, thus
determining the dependence of the deformed vierbein on the undeformed one.

The starting point is the assumption that space-time coordinates $x^{\mu}$ do
not commute
\[
\left[  x^{\mu},\,x^{\nu}\right]  =i\theta^{\mu\nu}%
\]
where $\theta^{\mu\nu}$ are assumed to be constant. However, under
diffeomorphism transformations, $\theta^{\mu\nu}$ becomes a function of $x$,
and one has to generalize the definition of the star product to be applicable
for a general manifold, but this is only known for symplectic manifolds
\cite{kont}. The effect of this noncommutativity is that ordinary products are
replaced with the star product defined by
\[
f\ast g=e^{\frac{i}{2}\theta^{\mu\nu}\frac{\partial}{\partial\xi^{\mu}}%
\frac{\partial}{\partial\eta^{\nu}}}f\left(  x+\xi\right)  g\left(
x+\eta\right)  |_{\xi=\eta=0}.
\]
In gauge theories one mainly uses $U(N)$ gauge fields subject to the condition
$\widehat{A}_{\mu}^{\dagger}=-\widehat{A}_{\mu}$ because such condition could
be maintained under the gauge transformations \cite{SW}
\[
\widehat{A}_{\mu}^{g}=\widehat{g}\ast\widehat{A}_{\mu}\ast\widehat{g}_{\ast
}^{-1}-\widehat{g}\ast\partial_{\mu}\widehat{g}_{\ast}^{-1},
\]
where $\widehat{g}\ast\widehat{g}_{\ast}^{-1}=1=\widehat{g}_{\ast}^{-1}%
\ast\widehat{g}$ . To avoid using complex or Hermitian gravitational fields,
we introduce the gauge fields $\widehat{\omega}_{\mu}^{\;AB}$ \cite{deformed}
subject to the conditions \cite{sheik},\cite{wess}
\begin{align*}
\widehat{\omega}_{\mu}^{\;AB\dagger}\left(  x,\theta\right)   &
=-\widehat{\omega}_{\mu}^{\;BA}\left(  x,\theta\right)  ,\\
\widehat{\omega}_{\mu}^{\;AB}\left(  x,\theta\right)  ^{r} &  \equiv
\widehat{\omega}_{\mu}^{\;AB}\left(  x,-\theta\right)  =-\widehat{\omega}%
_{\mu}^{\;BA}\left(  x,\theta\right)  .
\end{align*}
Expanding the gauge fields in powers of $\theta$, we have
\[
\widehat{\omega}_{\mu}^{\;AB}\left(  x,\theta\right)  =\omega_{\mu}%
^{\;AB}-i\theta^{\nu\rho}\omega_{\mu\nu\rho}^{\;AB}+\cdots.
\]
The above conditions then imply the following
\[
\omega_{\mu}^{\;AB}=-\omega_{\mu}^{\;BA},\quad\omega_{\mu\nu\rho}%
^{\;AB}=\omega_{\mu\nu\rho}^{\;BA}.
\]
A basic assumption to be made is that there are no new degrees of freedom
introduced by the new fields, and that these are related to the undeformed
fields by the Seiberg-Witten map \cite{SW}. This is defined by the property
\[
\widehat{\omega}_{\mu}^{\;AB}\left(  \omega\right)  +\delta_{\widehat{\lambda
}}\widehat{\omega}_{\mu}^{\;AB}\left(  \omega\right)  =\widehat{\omega}_{\mu
}^{\;AB}\left(  \omega+\delta_{\lambda}\omega\right)  ,
\]
where $\widehat{g}=e^{\widehat{\lambda}}$ and the infinitesimal transformation
of $\omega_{\mu}^{\;AB}$ is given by
\[
\delta_{\lambda}\omega_{\mu}^{\;AB}=\partial_{\mu}\lambda^{AB}+\omega_{\mu
}^{\;AC}\lambda^{CB}-\lambda^{AC}\omega_{\mu C}^{\;\;\;B},
\]
and for the deformed field it is
\[
\delta_{\widehat{\lambda}}\widehat{\omega}_{\mu}^{\;AB}=\partial_{\mu}%
\widehat{\lambda}^{AB}+\widehat{\omega}_{\mu}^{\;AC}\ast\widehat{\lambda}%
^{CB}-\widehat{\lambda}^{AC}\ast\widehat{\omega}_{\mu}^{\;CB}.
\]
To solve this equation we first write
\begin{align*}
\widehat{\omega}_{\mu}^{\;AB} &  =\omega_{\mu}^{\;AB}+\omega_{\mu}%
^{\prime\;AB}\left(  \omega\right)  \\
\widehat{\lambda}^{AB} &  =\lambda^{AB}+\lambda^{^{\prime}AB}\left(
\lambda,\omega\right)
\end{align*}
where $\omega_{\mu}^{\prime AB}\left(  \omega\right)  $ and $\lambda
^{^{\prime}AB}\left(  \lambda,\omega\right)  $ are functions of $\theta$, and
then substitute into the variational equation to get \cite{SW}
\begin{align*}
&  \omega_{\mu}^{\prime AB}\left(  \omega+\delta\omega\right)  -\omega_{\mu
}^{\prime AB}\left(  \omega\right)  \\
&  =\partial_{\mu}\lambda^{^{\prime}AB}+\omega_{\mu}^{\;AC}\lambda_{C}%
^{\prime\;B}-\lambda^{\prime AC}\omega_{\mu C}^{\;\;\;B}+\omega_{\mu}%
^{\prime\;AC}\lambda_{C}^{\;B}-\lambda^{AC}\omega_{\mu C}^{\prime\;\;B}\\
&  +\frac{i}{2}\theta^{\nu\rho}\left(  \partial_{\nu}\omega_{\mu}%
^{\;AC}\partial_{\rho}\lambda_{C}^{\;B}+\partial_{\rho}\lambda^{AC}%
\partial_{\nu}\omega_{\mu C}^{\;\;\;B}\right)
\end{align*}
This equation is solved, to first order in $\theta$, by
\begin{align*}
\widehat{\omega}_{\mu}^{\;AB} &  =\omega_{\mu}^{\;AB}-\frac{i}{4}\theta
^{\nu\rho}\left\{  \omega_{\nu},\,\partial_{\rho}\omega_{\mu}+R_{\rho\mu
}\right\}  ^{AB}+O(\theta^{2})\\
\widehat{\lambda}^{AB} &  =\lambda^{AB}+\frac{i}{4}\theta^{\nu\rho}\left\{
\partial_{\nu}\lambda,\,\omega_{\rho}\right\}  ^{AB}+O(\theta^{2})
\end{align*}
where we have defined the anticommutator $\left\{  \alpha,\,\beta\right\}
^{AB}\equiv\alpha^{AC}\beta_{C}^{\;B}+\beta^{AC}\alpha_{C}^{\;B}.$ With this
it is possible to derive the differential equation that govern the dependence
of the deformed fields on $\theta$ to all orders
\[
\delta\widehat{\omega}_{\mu}^{\;AB}\left(  \theta\right)  =-\frac{i}{4}%
\theta^{\nu\rho}\left\{  \widehat{\omega}_{\nu},_{\ast}\,\partial_{\rho
}\,\widehat{\omega}_{\mu}+\widehat{R}_{\rho\mu}\right\}  ^{AB}%
\]
with the products in the anticommutator given by the star product, and where
\[
\widehat{R}_{\mu\nu}^{\;\;AB}=\partial_{\mu}\widehat{\omega}_{\nu}%
^{\;AB}-\partial_{\nu}\widehat{\omega}_{\mu}^{\;AB}+\widehat{\omega}_{\mu
}^{\;AC}\ast\widehat{\omega}_{\nu C}^{\;\;B}-\widehat{\omega}_{\nu}^{\;AC}%
\ast\widehat{\omega}_{\mu}^{\;CB}%
\]

We are mainly interested in determining $\widehat{\omega}_{\mu}^{\;AB}\left(
\theta\right)  $ to second order in $\theta$. This is due to the fact that the
deformed gravitational action is required to be hermitian. The undefomed
fields being real, then implies that all odd powers of $\theta$ in the action
must vanish. The above equation could be solved iteratively, by inserting the
solution to first order in $\theta$ in the differential equation and
integrating it. The second order corrections in $\theta$ to $\widehat{\omega
}_{\mu}^{\;AB}$ are
\begin{align*}
&  \frac{1}{32}\theta^{\nu\rho}\theta^{\kappa\sigma}\left(  \left\{
\omega_{\kappa},\,2\left\{  R_{\sigma\nu},\,R_{\mu\rho}\right\}  -\left\{
\omega_{\nu},\left(  D_{\rho}R_{\sigma\mu}+\partial_{\rho}R_{\sigma\mu
}\right)  \,\right\}  -\partial_{\sigma}\left\{  \omega_{\nu,}\,\left(
\partial_{\rho}\omega_{\mu}+R_{\rho\mu}\right)  \right\}  \right\}
^{AB}\right.  \\
&  \qquad\qquad\quad\left.  +\left[  \partial_{\nu}\omega_{\kappa}%
,\,\partial_{\rho}\left(  \partial_{\sigma}\omega_{\mu}+R_{\sigma\mu}\right)
\right]  ^{AB}-\left\{  \left\{  \omega_{\nu},\,\left(  \partial_{\rho}%
\omega_{\kappa}+R_{\rho\kappa}\right)  \right\}  ,\,\left(  \partial_{\sigma
}\omega_{\mu}+R_{\sigma\mu}\right)  \right\}  ^{AB}\right)
\end{align*}
One problem remains of how to determine the dependence of the vierbein
$\widehat{e}_{\mu}^{a}$ on the undeformed field as it is not a gauge field. To
resolve this problem we adopt the strategy of considering the field $e_{\mu
}^{a}$ as the gauge field of the translation generator of the inhomegenious
Lorentz group, obtained through the contraction of the group $SO(4,1)$ to
$ISO(3,1).$ We write $\widehat{\omega}_{\mu}^{\;a5}=k\widehat{e}_{\mu}^{\;a}$
and $\widehat{\omega}_{\mu}^{\;55}=k\widehat{\phi}_{\mu}.$ We shall only
impose the condition $T_{\mu\nu}^{\;\;a}=0$ and not $\widehat{T}_{\mu\nu
}^{\;\;a}=0$ because we are not interested in $\phi_{\mu}$ which will drop out
in the limit $k\rightarrow0.$ The result for $\widehat{e}_{\mu}^{\;a}$ in the
limit $k\rightarrow0$ is
\begin{align*}
\widehat{e}_{\mu}^{\;a} &  =e_{\mu}^{a}-\frac{i}{4}\theta^{\nu\rho}\left(
\omega_{\nu}^{\;ac}\partial_{\rho}e_{\mu c}+\left(  \partial_{\rho}\omega
_{\mu}^{\;ac}+R_{\rho\mu}^{\;\;\;ac}\right)  e_{\nu c}\right)  \\
&  +\frac{1}{32}\theta^{\nu\rho}\theta^{\kappa\sigma}\left(  2\left\{
R_{\sigma\nu},R_{\mu\rho}\right\}  ^{ac}e_{\kappa c}-\omega_{\kappa}%
^{ac}\left(  D_{\rho}R_{\sigma\mu}^{\;\;\;cd}+\partial_{\rho}R_{\sigma\mu
}^{\;\;\;cd}\right)  e_{\nu d}\right.  \\
&  -\left\{  \omega_{\nu},\left(  D_{\rho}R_{\sigma\mu}+\partial_{\rho
}R_{\sigma\mu}\right)  \right\}  ^{ad}e_{\kappa d}-\partial_{\sigma}\left\{
\omega_{\nu},\,\left(  \partial_{\rho}\omega_{\mu}+R_{\rho\mu}\right)
\right\}  ^{ac}e_{\kappa c}\\
&  -\omega_{\kappa}^{\;ac}\partial_{\sigma}\left(  \omega_{\nu c}%
^{\;\;d}\partial_{\rho}e_{\mu d}+\left(  \partial_{\rho}\omega_{\mu}%
^{\;cd}+R_{\rho\mu}^{\;\;\;cd}\right)  e_{\nu d}\right)  +\partial_{\nu}%
\omega_{\kappa}^{\;ac}\partial_{\rho}\partial_{\sigma}e_{\mu c}\\
&  -\partial_{\rho}\left(  \partial_{\sigma}\omega_{\mu}^{\;ac}+R_{\sigma\mu
}^{\;\;\;ac}\right)  \partial_{\nu}e_{\kappa c}-\left\{  \omega_{\nu
},\,\left(  \partial_{\rho}\omega_{\kappa}+R_{\rho\kappa}\right)  \right\}
^{ac}\partial_{\sigma}e_{\mu c}\\
&  -\left.  \left(  \partial_{\sigma}\omega_{\mu}^{\;ac}+R_{\sigma\mu
}^{\;\;\;ac}\right)  \left(  \omega_{\nu c}^{\;\;d}\partial_{\rho}e_{\kappa
d}+\left(  \partial_{\rho}\omega_{\kappa c}^{\;\;d}+R_{\rho\kappa}%
^{\;\;\;cd}\right)  e_{\nu d}\right)  \right)  +O\left(  \theta^{3}\right)
\end{align*}

At this point, it is possible to determine the deformed curvature and use it
to calculate the deformed action given by \cite{deformed}
\[
\int d^{4}x\epsilon^{\mu\nu\rho\sigma}\epsilon_{abce}\widehat{e}_{\mu}^{a}%
\ast\widehat{R}_{\nu\rho}^{\;\;\;bc}\ast\widehat{e}_{\sigma}^{d}%
\]
Of course the actual expression obtained after substituting for the fields
$e_{a\rho\tau}^{\mu},$ $e_{a\rho\tau\kappa\sigma}^{\mu}$, $\omega_{\mu\rho
\tau}^{\;ab}$ and $\omega_{\mu\rho\tau\kappa\sigma}^{\;ab}$ is very
complicated, and it is not clear whether one can associate a geometric
structure with it. One can, however, take this expression and study the
deformations to the graviton propagator, which will receive $\theta^{2}$ corrections.

\section{Conclusions}

The simple idea that started with Utiyama to formulate the general theory of
relativity as a gauge theory of the Lorentz group has grown to become a
powerful tool in investigating gravitational theories. We have sampled only
few of the known applications in the literature, the full extent of which is
considerable. The main advantages are the simplicity and straightforwardness
of the formalism. The idea can be applied to shed light on the various aspects
of gravity. We have shown that a consistent formulation of massive gravity is
possible through the use of spontaneous breakdown of gauge symmetry. When
applied to graded Lie algebras it gives supergravity, and for the Chern-Simons
action it gives topological gravity. By extending the gauge algebra to become
complex, one obtains complex gravity with a Hermitian metric. It is also
possible to give a consistent deformation of the Einstein action on spaces
where ordinary products are replaced with star products. All this work gives
the promise that the gauge principle can unify gravity with the other
fundamental interactions, all of which are known to be based on the gauge theories.

\section{Acknowledgment}

This research is supported in part by the National Science Foundation under
Grant No. Phys-0313416.

\end{document}